\documentclass[prd,superscriptaddress,twocolumn,preprintnumbers,nofootinbib,endnote]{revtex4-1}
\pdfoutput=1

\usepackage{amsfonts,amssymb,amsmath}
\usepackage{bm}
\usepackage{bbm}
\usepackage{epsfig}
\usepackage{color}
\newcommand{\avg}[1]{\langle#1\rangle}

\newcommand{\SUC}{\text{SU(3)}_C}
\newcommand{\SUL}{\text{SU(2)}_L}
\newcommand{\UY}{\text{U(1)}_Y}
\newcommand{\Uem}{\text{U(1)}_\text{em}}
\newcommand{\UX}{\text{U(1)}_X}
\newcommand{\UR}{\text{U(1)}_\text{R}}

\newcommand{\jets}{\text{jets}}
\newcommand{\ETmiss}{\slashchar{E}_T}

\newcommand{\tev}{~\text{TeV}}
\newcommand{\gev}{~\text{GeV}}

\newcommand{\pb}{~\text{pb}}
\newcommand{\fb}{~\text{fb}}

\newcommand{ \slashchar }[1]{\setbox0=\hbox{$#1$}   % set a box for #1
   \dimen0=\wd0                                     % and get its size
   \setbox1=\hbox{/} \dimen1=\wd1                   % get size of /
   \ifdim\dimen0>\dimen1                            % #1 is bigger
      \rlap{\hbox to \dimen0{\hfil/\hfil}}          % so center / in box
      #1                                            % and print #1
   \else                                            % / is bigger
      \rlap{\hbox to \dimen1{\hfil$#1$\hfil}}       % so center #1
      /                                             % and print /
   \fi}                                             %

\begin{document}
\preprint{FERMILAB-PUB-11-067-T}

\title{The Gold-Plated Channel for Supersymmetric Higgs via Higgsphilic $Z'$ }

\author{Adam Martin}
\affiliation{Theoretical Physics Department, Fermilab, Batavia, IL 60510}

\author{Tuhin S. Roy}
\affiliation{Department of Physics, University of Washington, Seattle, WA
  98195-1560} 

\begin{abstract}

The lightest CP-even  Higgs boson in weak-scale supersymmetry can be discovered spectacularly early, even from $1\fb^{-1}$ of  data at  $7\tev$  LHC,  if it decays to a pair of light $Z'$, which in turn, decays to a pair of hard and ``isolated'' leptons. These $Z'$ must have infinitesimal couplings to light fermions in order to be consistent with precision electroweak constraints, while they have mild to moderate couplings to Higgs.   Hence  they are  {\em Higgsphilic}.  A $Z'$ with these properties appears at the electroweak scale in the ``viable'' gravity mediated supersymmetry breaking.   We construct an effective model to extract the $Z'$ phenomenology.  Even in a decoupled limit where all gauginos and sfermions are heavy and supersymmetry production is purely electroweak,  we find that
the Higgs boson as well as supersymmetry can be found early through the discovery of  $Z'$ in samples of events with $4\,\ell$ and  $4\,\ell + \ETmiss$ respectively.  Additionally, in cases where the $Z'$ is long-lived, we show that the trigger menus employed at the ATLAS detector to find long lived particles are capable of finding thousands of Higgs events from $1\fb^{-1}$ of  data. 

\end{abstract}

\maketitle

\section{Introduction}

The Higgs boson decaying to four hard and isolated leptons provides the cleanest channel (the so-called  ``gold-plated'' channel) for its discovery. The only significant  standard model (SM) background is di-boson ($ZZ$) production. Other, usually noxious SM backgrounds, such as $t\bar{t}$, $W/Z+\text{ jets}$, are  negligible. However, the gold-plated channel gets activated only for a relatively heavy SM Higgs boson ($m_h \gtrsim 130\gev$~\cite{Aad:2009wy}), when the Higgs can decay to four isolated leptons via $ZZ^*$ or $ZZ$.  Unfortunately, in weak-scale supersymmetry, which remains the most well-motivated ultraviolet (UV) extension of the standard model (SM), the mass of the lightest CP-even Higgs boson is predicted to be $\lesssim 125\gev$ for stop masses and mixings that do not exceed $1\tev$~\cite{Carena:2000dp} . Higgses in this mass range decay dominantly to $\bar{b} b$ and are notoriously hard to find.  Only by using sophisticated techniques such as jet-substructure  can we hope to find the Higgs in this mass range at $5~\sigma$ confidence from $10\fb^{-1}$ of data~\cite{Kribs:2009yh,Kribs:2010hp,Kribs:2010ii}. 

The goal of this paper is to open up the $4\,\ell$ mode for light, supersymmetric Higgses. This requires extending the Minimal Supersymmetric Standard Model (MSSM).  However, as we show, when the $4\,\ell$ channel is resurrected for light Higgses, the implications are spectacular.  A tiny branching fraction of $h \rightarrow 4\,\ell$ can render the Higgs easily discoverable, even with $1\fb^{-1}$ of data at the LHC operating at $\sqrt{s} = 7\tev$ center of mass energy.  

In many extensions of the MSSM, a pseudo-scalar (namely, $a$) can be lighter than the Higgs boson, and will mediate unconventional and exciting decay modes~\cite{Chang:2005ht, Carena:2007jk, Chang:2008cw}. The simplest unconventional mode is a light Higgs decaying to four final state particles, though decays to even higher multiplicity states are also possible~\cite{Chang:2005ht}. However, $a$ has Higgs-like couplings, so it decays dominantly to the heaviest particles that are kinematically allowed rather than to light leptons. Typically, $a \rightarrow \bar b b$, implying $h \rightarrow 2b\, 2\bar b$, but there are corners of parameter space where $2 m_\tau < m_a < 2 m_b $ and $a$ decays to $2\,b$ are forbidden. In these corners, $h \rightarrow 2 a \rightarrow 2 \tau\, 2 \mu$ has been shown to be a viable discovery mode~\cite{Lisanti:2009uy}. However, such ultra-light $a$ give rise to soft and, often, collinear muons, and are tricky to analyze.  

In numerous recent models of dark matter~\cite{ArkaniHamed:2008qn,Cheung:2009qd, Morrissey:2009ur}, as in hidden valley models~\cite{Strassler:2006im,Strassler:2006qa, Han:2007ae}, a new abelian gauge boson $Z'$ is introduced. If the Higgs couples to these $Z'$, $h \rightarrow Z'Z' \rightarrow 4\, \ell$ is feasible. Unfortunately, $Z'$ introduced for dark matter purposes tend to be extremely light $M_{Z'} \sim\gev$, making their decay products predominantly collinear and/or soft.  The final states from $h \rightarrow Z'Z' \rightarrow 4\, \ell$ do contain $4$ leptons, but rarely $4$ {\it hard} and {\it isolated} leptons.  Though studies have shown that it is possible to extract the Higgs by looking for jets made entirely of collinear leptons (so-called lepton jets~\cite{Falkowski:2010gv}), these methods are highly sensitive to detector effects. It remains to be seen how well these objects can be identified and how well the Higgs mass can be reconstructed in a realistic detector simulation. 

The introduction of a heavier Higgsphilic $Z'$  resurrects the gold-plated channel for discovery of a light Higgs. Specifically, this $Z'$ must be sufficiently heavy so that it remains un-boosted after the decay $h \rightarrow Z'Z'$.  Provided $M_{Z'} \gtrsim 0.1\,m_h$, the boost is sufficiently small and, consequently, massless $Z'$ decay products are separated enough to be individually distinguished in a detector.  An upper limit on the $Z'$ mass  comes simply from kinematics, $M_{Z'} \lesssim m_h/2$. In short, a $Z'$ mass on the order of the electroweak scale does the job quite well.  A hidden valley with such a $Z'$ was recently proposed for the sake of  involving the SM light Higgs to four fermions decay~\cite{Gopalakrishna:2008dv}.  However, the structure of supersymmetry severely restricts couplings and arbitrary mixing of a generic hidden abelian sector with Higgs is not allowed. 

The fusion of supersymmetry with Higgsphilic $Z'$ can do more than just facilitate Higgs discovery. In fact, combining multiple leptons signals with large missing energy, the result is an extremely clean channel for discovering supersymmetry itself. In the MSSM, multi-lepton plus $\slashchar E_T$ signals arise via cascades containing di-bosons or combinations of light sleptons and $W/Z$s. These rates are usually small, suppressed by the small fraction of cascades which contain $W/Z$. However, if a weak-scale $Z'$ is added to the model, we open up the possibility of large rates to $4\,\ell + \slashchar E_T$.  In this scenario, heavier neutralinos prefer to decay to lighter neutralinos  by emitting a $Z'$. As all supersymmetric events must terminate in a pair of the lightest neutralinos (the lightest supersymmetric particle in this setup), a large fraction of all supersymmetric events contain $Z'Z' + \slashchar E_T$, paving the way for many events with $4\,\ell + \slashchar E_T$. As we will show, the success of this channel is controlled essentially by the Higgsino masses. If the Higgsinos are light enough, the $4\, \ell + \slashchar E_T$ channel can bring discovery regardless of how heavy the  squarks, sleptons, and gluinos are.

Remarkably, a $Z'$ with exactly the desired properties is predicted in the ``viable'' gravity-mediated supersymmetry breaking model~\cite{Kribs:2010md}. In this model, the flavor problem~\cite{Gabbiani:1988rb,Gabbiani:1996hi,Bagger:1997gg,Ciuchini:1998ix}   associated with gravity mediation~\cite{Chamseddine:1982jx,Barbieri:1982eh,Ibanez:1982ee,Hall:1983iz,Ohta:1982wn,Ellis:1982wr,AlvarezGaume:1983gj,Nilles:1983ge,Nath:1983fp} is solved -- not by imposing continuous flavor symmetries~\cite{Barbieri:1995uv,Barbieri:1996ww} or 
gauged discrete family symmetries~\cite{Kaplan:1993ej,ArkaniHamed:1996xm}  to align the mass matrices~\cite{Hall:1990ac,Nir:1993mx}  -- 
but by the emergence of an approximate and an accidental $\UR$ symmetry. The complete model contains an additional gauged abelian symmetry (say, $\UX$),  which is broken along with the electroweak symmetry right at the electroweak scale.  Unlike the hidden sector models, the $\UX$  is completely {\it visible} to us since there exists electroweak doublets that carry $X-$charges.  The coexistence of supersymmetry along with an  electroweak scale Higgs-phylllic  $Z'$, ensures significant discovery potential for Higgs/higgsionos via the discovery of  $Z'$.  

Instead of analyzing the full model in Ref.~\cite{Kribs:2010md},  we construct a ``bare-bones'' model where we integrate out all particles not necessary for a discussion of $Z'$ phenomenology. The bare-bones model contains only the SM fermions, the CP-even lightest Higgs boson, Higgsinos, and an additional singlino.  This simplified approach to models has recently been advocated in Ref.~\cite{Alwall:2008ag}. 

Within current experimental constraints as calculated from the Tevatron results, we find that $Z'$ in the bare-bones model can be discovered easily in a sample of Higgs events (events with $4\,\ell$) as well as in a sample of supersymmetric events ($4\,\ell + \ETmiss$ events) easily and, thus, can pave the way for the discovery of supersymmetry and a supersymmetric Higgs. 

We begin with a description of the bare-bones model in Sec.~\ref{sec:model}. In Sec.~\ref{sec:pheno} we
show that the Tevatron data puts severe constraints on $h \rightarrow Z'Z' \rightarrow 4\, \ell$ branching fraction, and  on Higgsino masses. In the same section we also discuss  the discovery potential of $Z'$, Higgs, and Higgsinos at the LHC within these constraints.  For readers interested in how the bare-bones model arises from a more fundamental theory, we show the mapping between the viable gravity mediation scenario and the bare-bones model  in Sec.~\ref{sec:viable-grav-medi}.  Our concluding reflections are in Sec.~\ref{sec:conclusion}.

\section{The bare-bones model of a supersymmetric Higgsphilic $Z'$}
\label{sec:model}

We seek a bare-bones model with the signal
\begin{itemize}
\item $4\,\ell$: signature of Higgs decaying to $2 Z' \rightarrow 4\,\ell$. 
\item Inclusive $4\,\ell + \ETmiss$: signature of supersymmetric events with  $2Z' + \text{LSPs} \rightarrow 4\,\ell + \ETmiss$.
\end{itemize}
The bare-bones model is not manifestly supersymmetric, but it should be understood as an effective 
theory of a fully supersymmetric and UV-complete model. More precisely, it is the effective theory of the viable gravity mediation. The particle content of the bare-bones model is listed below:
\begin{table}[h]
\label{bare-bones-model-content}
\begin{tabular}{ c | c }
\hline
$h$          &  The lightest CP-even Higgs boson    \\
$Z'$         &  \parbox[c][0.4 in]{2.2in}{  
          New neutral gauge boson of a broken symmetry $\UX$, with mass  $M_{Z'}$ }\\ 
$\tilde S$   & \parbox[c][0.5 in]{2.2in}{ A Dirac singlino and the candidate for the lightest supersymmeric particles (LSP)}    \\ 
$\tilde H \equiv
\begin{pmatrix}
  \tilde C \\ \tilde N 
\end{pmatrix}
$  & \parbox[c][0.5 in]{2.2in}{ An electroweak doublet of Dirac inos. $\tilde N $ and $\tilde C $  are the neutral and charged components respectively.} \\
\hline
\end{tabular} 
\caption{List of particles in the bare-bones model}
\end{table}

Next, we compile the list of interactions involving $Z'$, that are necessary to generate the signals itemized in the beginning of this section.  
\begin{gather}
 -\frac{\epsilon}{2} \ Z'_{\mu \nu} B^{\mu \nu} 
 \label{kin-mix}\\
   g_0  \: M_{Z'}  \ h\, Z'_{\mu} Z'^{\mu} 
 \label{int-hz'z'}\\
   g_X\: \eta \ \bar{\tilde N} \gamma_\mu  P_L  \tilde S  \ Z'^{\mu}  
         \ + \ g_X \: \zeta \  \bar{\tilde N} \gamma_\mu P_R \tilde S  \  Z'^{\mu}  + H.c.
 \label{int-nhz'}
\end{gather}
In the above, $B_\mu$ is the  $\UY$ gauge boson, $\epsilon, g_0, \eta, \zeta$ are small coupling constants, and $g_X$ is the strength of broken abelian symmetry $\UX$. In the bare-bones model $g_X$ is weaker than the weak scale gauge couplings $g_Z$ or $g_W$, the strength of interactions between $Z$ or $W$ and their respective current. 

The coupling $\epsilon$  in Eq.~\eqref{kin-mix}  dictates the amount of mixing between $Z'$ and $B$. Its effects can be unearthed after redefining $B_\mu \rightarrow B_\mu - \epsilon Z'_\mu$. Such a redefinition, undoes the mixing  but generates coupling between the hypercharge current $J_Y$ and $Z'$.  Consequently, non zero Higgs vacuum expectation value (vev)  gives rise to an additional $Z-Z'$ mixing. Undoing the $Z-Z'$ mixing, we find an additional coupling between $J_Z$ (the $Z-$current) and $Z'$~\cite{Gopalakrishna:2008dv}. The effect of the operator in Eq.~\eqref{kin-mix} can be summarized in a following interaction that shows the couplings of $Z'$ with the SM fermions in a straight-forward way:   
\begin{equation}
  \label{eq:kin-mix-final}
      \epsilon  \:  g _Y  Z'_\mu  \left(    J_Y^\mu \ + \  
             \frac{M_Z^2}{M_Z^2 - M_{Z'}^2}  J_Z^\mu \right) \   \; ,
\end{equation}
where $g_Y$ is the hypercharge gauge coupling. Since the parameter $\epsilon$ is constrained to be less than or in the order of $10^{-2}$~\cite{Hook:2010tw}, as long as $Z'$ appears at the electroweak scale, the couplings in Eq.~\eqref{eq:kin-mix-final} have little phenomenological impact. However, the $\mathcal O(\epsilon)$ couplings do become important if the $Z'$ is kinematically restricted and can only decay via Eq.~\eqref{eq:kin-mix-final}. In this case, while the total width of $Z'$ depends on $\epsilon$, its branching fractions to various SM fermions do not. In fact, in the limit $M_Z^2 \gg M_{Z'}^2$,  the branching fractions are  completely determined:
\begin{equation}
  \label{eq:5}
  \begin{split}
  & \text{Br}\left( Z' \rightarrow \tau^+ \tau^- \right) \sim 13\%  \quad
   \text{Br}\left( Z' \rightarrow \ell^+ \ell^- \right) \sim 28\%   \\
   & \quad  \text{Br}\left( Z' \rightarrow q \bar{q} \right) \sim 60\%  \quad \quad
   \text{Br}\left( Z' \rightarrow b \bar{b} \right) \sim 7\%  \\
  & \qquad \qquad \qquad \quad
          \text{Br}\left( Z' \rightarrow \nu \nu \right) \lesssim 1\%   \; ,
  \end{split}
\end{equation}
where  $q$  refers to all quarks and $\nu$ refers to neutrinos of all flavor.

Since the bare-bones model has been pointed out to be an effective description of a supersymmetric UV complete model, the Higgs boson $h$ should be considered to be the part of a fully supersymmetric Higgs sector.   We will show later in Sec.~\ref{sec:viable-grav-medi} that when the bare-bones model is derived from the viable gravity mediation,  the couplings of the Higgs to the SM fermions and the electroweak gauge bosons remain SM-like to the leading order of small model parameters.  Higgs in the bare-bones  inherits all SM interactions with approximately SM-like strengths. 
Its production is thus identical to a SM Higgs.  The only difference in properties of the Higgs boson is that  if $m_h > 2 M_{Z'}$, it can decay to a pair of $Z'$ because of the operator in  Eq.~\eqref{int-hz'z'}. The branching width of $h \rightarrow 2 Z' \rightarrow 4\,\ell$ is quadratically sensitive to the coupling constant $g_0$. However, because of the tiny total decay width of the lightest CP-even MSSM Higgs, even a small $g_0$ results in a large branching fraction.  

Operators in Eq.~\eqref{int-nhz'} are also responsible for the decay of  heavier neutralinos to lighter neutralinos with the emission of a $Z'$. Note that, since $\tilde N$ carries electroweak charge, it can be produced by Drell-Yan processes at the LHC, while $\tilde S$ cannot. Hence, realistically speaking, we must also assume $M_{\tilde N} > M_{\tilde S}$ in order to generate any $Z'$ events. 
To flesh out what the bare-bones model predicts for supersymmetric events,  we need to list additional interactions. These interactions do not involve $Z'$ directly, but have important implications for $Z'$ phenomenology. 
\begin{gather}
   - \frac{g_Z}{2}\: \eta \ \bar{\tilde N} \gamma_\mu  P_L  \tilde S   \ Z^{\mu}  
         \ - \ \frac{g_Z}{2} \: \zeta \  \bar{\tilde N} \gamma_\mu P_R \tilde S   \ Z^{\mu} + H.c.
\label{int-nhz}  \\
   - \frac{g_Z}{\sqrt{2}}\: \eta \ \bar{\tilde C} \gamma_\mu  P_L  \tilde S   \ W^{\mu}  
         \ - \ \frac{g_Z}{\sqrt{2}} \: \zeta \  \bar{\tilde C} \gamma_\mu P_R \tilde S   \ W^{\mu}+H.c.
 \label{int-ncw}
\end{gather}
The operators in Eq.~\eqref{int-nhz} allow $\tilde N \rightarrow \tilde S + Z$ decay, however as long as $ M_{Z} > ( M_{\tilde N}  - M_{\tilde S} ) > M_{Z'}$,  the decay mode $\tilde N \rightarrow \tilde S + Z' $ dominates. For other mass hierarchies, both the decay modes compete. Decays to $Z$ are slightly suppressed because of a larger $Z$ mass, but are enhanced in the small $g_X/g_Z$ limit.

The last set of operators (Eq.~\eqref{int-ncw}) provide an alternative path for charged Higgsinos ({\it i.e.} $\tilde C$) to decay to neutralinos, $\tilde C \rightarrow \tilde S + W$. As $\tilde C$ and $\tilde N$ are part of the same electroweak doublet, they are much closer in mass than $\tilde C$ and $\tilde S$. Consequently, kinematics suppresses $\tilde C\rightarrow \tilde N + W$ compared to $\tilde C\rightarrow \tilde S + W$, with the $W$ in the former case often off-shell. Thus, neither chargino pair production nor associated chargino plus neutralino production will generate events contains $2\,Z'$; chargino pairs lead to $W^+W^- + \slashchar E_T$, while chargino plus neutralino leads to  $W^{\pm}Z' + \slashchar E_T$. As a result, the only supersymmetric source of $4\,\ell + \slashchar E_T$ events is neutralino pair production. \\

Summarizing, 
\begin{itemize}
\item  The $Z'$ branching fractions to various SM fermions is completely determined given its mass. 
It has a large width to leptons and, hence, is easier to find in $Z' \rightarrow 2 \ell$. Unlike $Z$ bosons, the $Z'$ width to neutrinos is minuscule, suppressed by $\left( M_{Z'}/ M_Z \right)^2$ compared to the leptonic width.
\item  The channel $h \rightarrow Z'Z' \rightarrow 4\,\ell$ is a clean channel for the simultaneous discovery of $Z'$ and Higgs as long as $m_h > 2 M_{Z'}$. For a given $m_h$ and $M_{Z'}$, the rates of $4\,\ell$ events depend only on the branching fraction of $h \rightarrow Z'Z'$, which is governed by the parameter $g_0$.  
\item  Inclusive $4\,\ell + \ETmiss$ is another clean signature to find $Z'$ along with supersymmetry in the bare-bones model. Provided $ M_{Z} > ( M_{\tilde N}  - M_{\tilde S} ) > M_{Z'}$ almost all neutralino-pair events contain two $Z'$, and the rate for $\tilde N \tilde N \rightarrow 2 Z' + 2 \tilde S \rightarrow 4 \ell + \ETmiss$  depends only on the mass of $\tilde N$. 
\end{itemize}

\section{New Physics via Signals of $Z'$ at the Collider}
\label{sec:pheno}

In this section, we investigate the discovery potential of three distinct signals of the bare-bones model:  $(i)$ signals of Higgs in $4\,\ell$, $(ii)$ signals of supersymmetry in $4\,\ell  + \ETmiss$, and finally  
$(iii)$ signals of long lived $Z'$ in  $ h \rightarrow Z'Z'$ events. 

\subsection{ $4 \ell$ without Missing Energy}
\label{sec:higgs_prompt}

Unlike most collider studies of the lightest CP-even Higgs boson in the MSSM~\cite{Djouadi:2005gj}, we seek Higgses produced via gluon fusion.      
\begin{equation}
  \label{eq:h-topo}
  p p \text{ or } p \bar{p} \rightarrow h \rightarrow Z' Z'  \rightarrow 4\,\ell \; .
\end{equation}
In the narrow-width approximation, this rate depends on the product of the production cross-section and the partial decay width of Higgs to $Z'$. As mentioned before, Higgses in the bare-bones model carries all the usual SM interactions with approximately SM-like strength couplings. Hence, the production cross-section of the bare-bones Higgs is essentially the same as the SM Higgs production cross section,
\begin{equation}
  \label{eq:h-prod}
 \begin{split}
  \sigma \left( gg \rightarrow h \right) \big|_\text{bare-bones} 
             \  
              \simeq \ &
                      \sigma \left( gg \rightarrow h \right) \big|_\text{SM}  \; .
 \end{split}
\end{equation}

The partial decay width of Higgs to $Z'$ depends on the coupling $g_0$ in Eq.~\eqref{int-hz'z'}.  The strongest constraint on the decay width, and hence on $g_{0}$, comes from the Tevatron. Even though CDF and D0 are not currently looking for  $h \rightarrow 4\, \ell$ for a light Higgs, both experiments do have inclusive and isolated $4\,\ell$ searches, primarily to measure the $ZZ$ cross-section at the Tevatron.  The CDF search at $5\,\fb^{-1}$ is public~\cite{CDF:10238}.  They see $6$ candidate events with $4$ isolated leptons, among which $4$ events are selected to be good candidates for the decay product of $ZZ$. One of the remaining events fits the profile of a $ZZ\rightarrow 2\, \ell + 2\, \tau$ event, where both the $\tau$s have decayed leptonically. The last remaining event is {\it curious}, since at first glance it does appear to have stemmed from a couple of promptly-decaying  resonances with a mass of $\sim\,55\gev$.  Although a single event can by explained away by a possible mis-recombination of leptons, it remains to be seen whether the early runs of LHC and even CDF/D0 data at larger luminosity  finds similar anomalous events in the same mass-window. 

In this work, we impose bounds on $\sigma_h \times \text{Br}\left( h \rightarrow Z'Z' \right)$ based on null observation of candidate $4\, \ell$ events. The maximum number of allowed $4\, \ell$ events at $95\%$ confidence level ($3.0$) is translated into a cross section bound by equating the $ZZ$ cross section measured by CDF ($1.7^{+1.2}_{-0.7}\, \pm 0.2\pb$) with four $4\, \ell$ events~\cite{CDF:10238}.  At $95\%$ confidence level
\begin{equation}
  \label{eq:cdf-lim-br}
\begin{split}
  \sigma_h \times \text{Br}\left( h \rightarrow Z'Z' \right) \leq  \frac{3}{4} \:
    \sigma_{ZZ} \:  \frac{\text{Br}(Z \rightarrow 2\ell)^2 }{\text{Br}(Z' \rightarrow 2\ell)^2 }\,
      \mathcal A_\text{rel},
\end{split}
\end{equation}
where we have divided by the square of the $Z' \rightarrow \ell^+\ell^-$ branching fraction of Eq.~\eqref{eq:5}%
\footnote{This procedure assumes that the Higgs width is small. However, for the range of Higgs masses and parameters we are considering this is a perfectly valid approximation.}. The factor $\mathcal A_{rel}$ contains the ratio of acceptances for $ZZ$ events compared to $Z'Z'$ events. We divide this ratio into three factors:
\begin{equation}
\mathcal A_\text{rel} = a_\text{fiducial} \times a_{4\,\ell-\text{cut}} \times a_\text{id}.
\label{eq:effs}
\end{equation}
The $a_\text{fiducial}$ factor designates how often all four leptons from $ZZ \rightarrow 4\,\ell$ events pass CDF's tight kinematic criteria ($p_T > 10\gev, \  |\eta| < 1.1, \text{ and } \Delta R_{\ell\ell} > 0.4$) compared to leptonic $Z'Z'$ events.  As the $Z$ and $Z'$ have different masses and different production mechanisms, there is  slight difference in kinematic acceptance. The second factor $ a_{4\,\ell-\text{cut}} $ is the ratio of acceptance for the total lepton invariant mass cut, $m_{4\,\ell} < 300\gev$%
\footnote{ $m^2_{4\,\ell} = (\sum_{i=1}^4 p_{\ell_i})^2\; .$}
, while the final factor $a_\text{id}$ accounts for any difference in identifying/reconstructing leptons ({\em after} passing all kinematic cuts) related to the origin of the lepton.  We take $a_{id}$ to be $1$ in this work. 

The total SM $ZZ \rightarrow 4\,\ell$ acceptance can be found in Ref.~\cite{CDF:10238}. However, the quoted fraction ($\sim 0.11$) contains the id efficiency, which we have just argued should cancel in a ratio such as Eq.~\eqref{eq:effs}. Therefore, to estimate $a_\text{fiducial}$ and $a_{4\,\ell-\text{cut}}$, we resort to Monte Carlo. We generate samples of  $p\bar p \rightarrow ZZ \rightarrow 4\,\ell$ and $p \bar p \rightarrow h \rightarrow Z'Z' \rightarrow 4 \ell$ with MadGraphV4~\cite{Alwall:2007st} + PYTHIA6.4~\cite{Sjostrand:2006za}. These events are passed through the fast-detector simulator PGS~\cite{PGS}, then analyzed. Following this procedure, we find a kinematic acceptance of $\sim 0.2$ for $ZZ$ and $\sim 0.15$ for $h\rightarrow Z'Z'$. In the surviving events we next form the total lepton invariant mass and impose the cut $m_{4\,\ell} < 300\gev$. Roughly $\sim 86\% $ of SM events pass this cut, while -- given that the four leptons sum to $m_h$ in the $Z'$ case, all $Z'$ events pass. The Higgs mass used in these samples was $120\gev$, though we expect identical  numbers for all $m_h \lesssim 125\gev$\@%
\footnote{ Numbers produced above assume relatively heavy $Z'$. For $M_{Z'} \lesssim 15\gev$, a significant numbers of  decayed leptons are not isolated. As a  consequence, the acceptance of $Z'Z'$ events decrease drastically with decreasing $M_{Z'}$.}.

\begin{figure*}[!ht]
\includegraphics[height=2.75in]{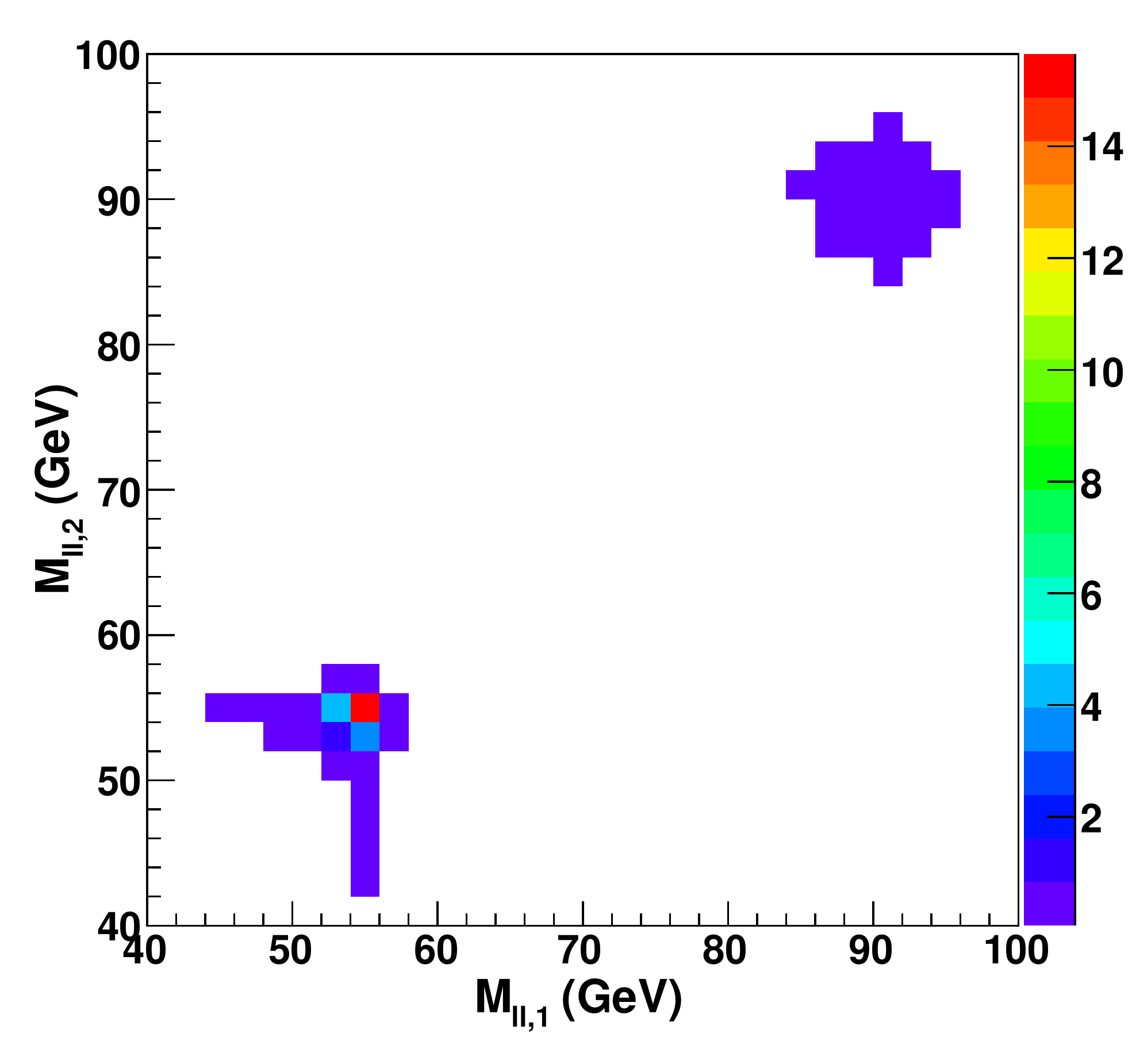}
\hspace{0.5cm}
\includegraphics[height=2.75in]{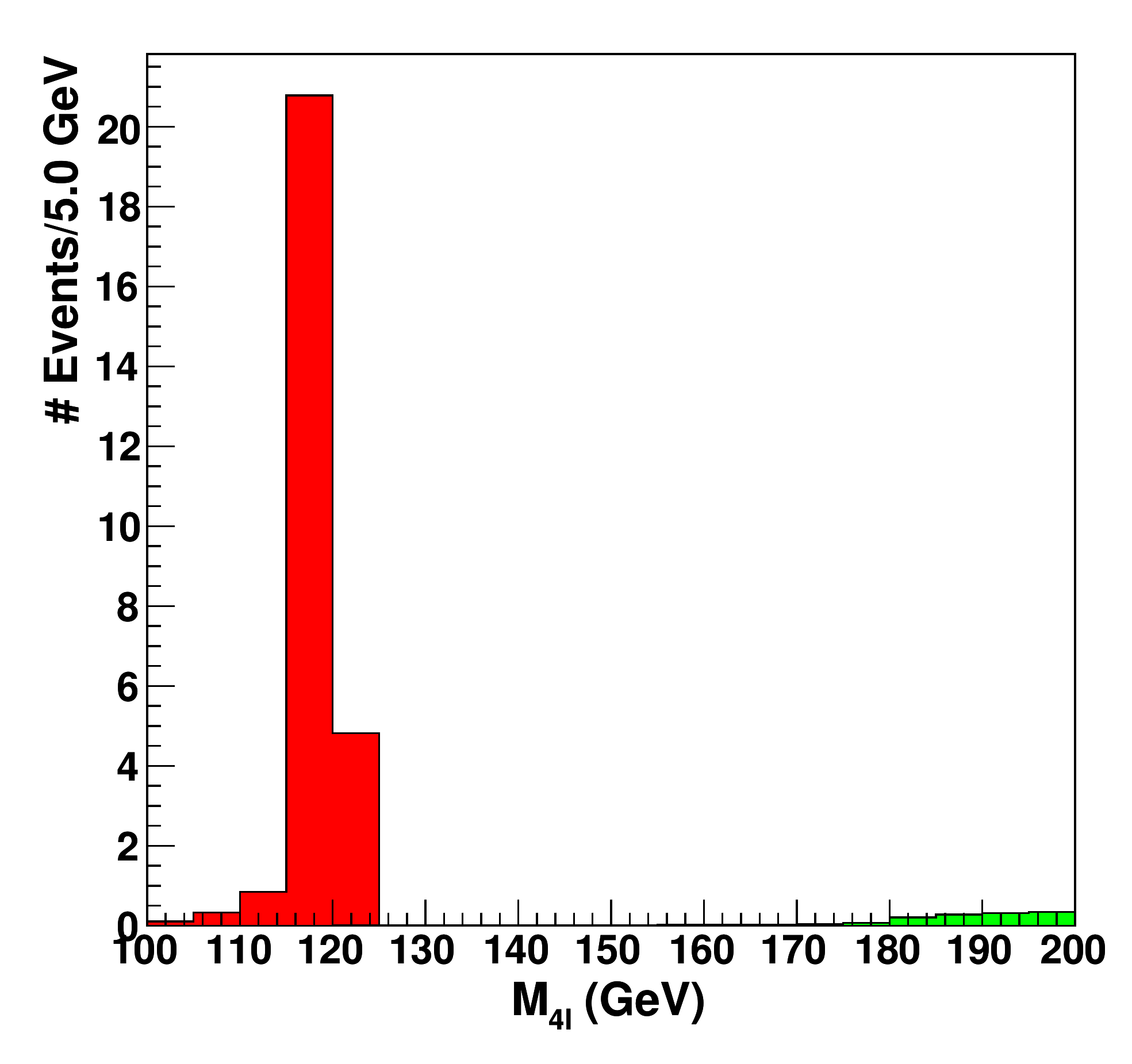}
\caption{In the left pane we display LHC $h \rightarrow Z'Z' \rightarrow 4\,\ell$ events in a two-dimensional plane $M_{\ell\ell,1}, M_{\ell\ell,2}$ where $M_{\ell\ell,1}$ ($M_{\ell\ell,2}$) is the lepton pair with highest $p_T$ (subleading $p_T$), and we stack signal and background events on top of each other. The cluster of events at $\sim 55\gev$ shows the prominence of the $Z'$ signal (color online). For the same events, we plot the total $4$ lepton invariant mass in the right-hand pane. Both signal (red) and background (green) events have been generated with MadGraph. This figure is analogous to Figure.~12 in Ref~\cite{CDF:10238} created for $4\,\ell$ events at the Tevatron. }
\label{fig:fourlep}
\end{figure*}

Combining these factors, our derived value of $\mathcal A_\text{rel}$ is $1.15$. Plugging in to Eq.~\eqref{eq:cdf-lim-br}, we find that 
\begin{equation}
  \label{eq:limit-h-cs}
  \sigma_h \times \text{Br}\left( h \rightarrow Z'Z' \right)  \ \lesssim \ 84\fb \quad @95\% \text{CL}
\end{equation}
Dividing by the Higgs production cross section at $\sqrt s = 1.96\tev$ (at NNLO~\cite{Anastasiou:2008tj}), we translate the cross section bound into a (Higgs mass dependent) limit on $\text{BR}(h \rightarrow Z'Z')$:
\begin{equation}
\text{BR}(h \rightarrow Z'Z') \leq \  7.7\times 10^{-2} \quad  \text{for } m_h = 120\gev\; .
 \label{eq:brlimits}
\end{equation}

We can immediately turn these branching fraction limits into predictions for the LHC. We simply reverse the process used to obtain $BR(h \rightarrow Z'Z')$, replacing the Tevatron Higgs production cross section with the most recent calculation of $\sigma(pp \rightarrow h)$ at $\sqrt s = 7\tev$~\cite{Dittmaier:2011ti}. While this gives us the net number of four-lepton events, we still need to account for acceptance. We estimate the acceptance using Monte-Carlo events, generated as before. Events are kept if they contain four isolated leptons with $p_T > 10\gev, |\eta| < 2.5$, $\Delta R_{\ell\ell} > 0.4$.  We check further that, out of these $4\,\ell$ events, there are two pairs of same-flavor, opposite sign leptons.  After this procedure, we find that roughly $\epsilon_{\ell} \sim 28\%$  of signal four-lepton events survive. Combining these factors, we find:
\begin{equation}
  \label{eq:lhc-h}
  \begin{split}
    \sigma(pp \rightarrow Z'Z' \rightarrow 4\,\ell) _{\text{LHC}@7}  & \simeq  \\
        \sigma(pp \rightarrow h)_{\text{LHC}@7}  \times \text{BR}  &
            \left( h  \rightarrow Z'Z' \rightarrow 4\,\ell\right) \times \epsilon_{\ell} \\
              & \lesssim  \ 28~\text{events }  \fb^{-1} \; .
  \end{split}
\end{equation}

The  $Z'Z' \rightarrow 4\,\ell$ events have essentially no standard model background. Di-boson production,  $pp \rightarrow ZZ $ followed by $ZZ \rightarrow 4\,\ell$, is the only SM process that results in $4$ high-$p_T$ leptons produced in the primary interaction. Of the di-boson backgrounds, $Z\rightarrow e^+ e^-$ and $Z\rightarrow \mu^+\mu^-$ can be easily recognized by looking at the invariant mass of the lepton pair, $M_{\ell\ell}$. Provided $M_{Z'} \ne M_Z$, these background events will be well-separated from the signal. Leptonically decaying $\tau^{\pm}$ pairs will have a continuous spectrum in $M_{\ell\ell}$ and could be problematic. However, the rate for this process is additionally suppressed by leptonic branching fractions of the $\tau$ and by cut acceptance. Each of the $\tau$ daughters has only a fraction of the $\tau$ momentum, so daughter leptons are less  likely to pass the basic kinematic cuts than $e/\mu$ coming directly from the $Z$.  
%Further, we do not expect them to show up in same place as the signal events in the   $(M_{\ell\ell}-M_{\ell\ell})$ plane. 

Other, reducible backgrounds in this channel come from the mis-identification of jets as leptons. Using the detector simulation PGS to model lepton-misidentification, we have checked the contributions in the $4\,\ell$ final state from $Z(\ell\ell) + \jets$, $WW/WZ$, and leptonic $t\bar t + \jets$. We find that none of these leave a trace in the signal region. Given that the probability for a jet to fake a lepton is, very conservatively, $0.1\%$, the small size of the reducible background is not completely surprising. 

Combining the backgrounds with the signal, we show all $4\,\ell$ events in Fig.~\ref{fig:fourlep}. To generate the signal, we use a sample  point, $m_h = 120\gev, M_{Z'} = 55\gev$, and $BR(h\rightarrow Z'Z') = 7.7\%$. We first plot events in the  $M_{\ell\ell,1}-M_{\ell\ell,2}$ plane, where $M_{\ell\ell,1} \text{ and }M_{\ell\ell,2}$ correspond to the mass of the lepton pair with leading  and subleading $p_T$ respectively. 
Fig.~\ref{fig:fourlep}  clearly shows the cleanliness of the $Z'$ signal. For  our choice of $Z'$ mass the leptonic $Z$ background events are well separated from the signal and the total $4\,\ell$ invariant mass for each event finds the Higgs peak cleanly. 
 
Both the Higgs and $Z'$ signals are extremely clean and clearly spectacular. By optimizing the lepton acceptance (by, for example, admitting lower $p_T$), it may be possible to increase the rate even further. While the signal above shows a $h\rightarrow Z'Z'$ branching fraction right on the edge of the Tevatron bound, slightly lower branching fractions would still lead to shockingly early light Higgs discovery. Even with a $h\rightarrow Z'Z'$ branching fraction as low as $1.4\%$, we expect $\sim 5$ signal events within $1\fb^{-1}$ of $7\tev$ LHC running -- discover far sooner than with traditional light-Higgs modes like $\gamma\gamma$ or $\tau^+\tau^-$. Note that the signals and rates shown in this section apply to promptly decaying $Z'$ only. For Higgses decaying to long-lived $Z'$s, a completely different set of signals, equally spectacular to those in the prompt case, are possible. We explore this case later in this section. 

\subsection{$4\,\ell$ with Missing Energy }
\label{sec:higgsino-pheno}

\begin{figure*}
\includegraphics[height=2.75in]{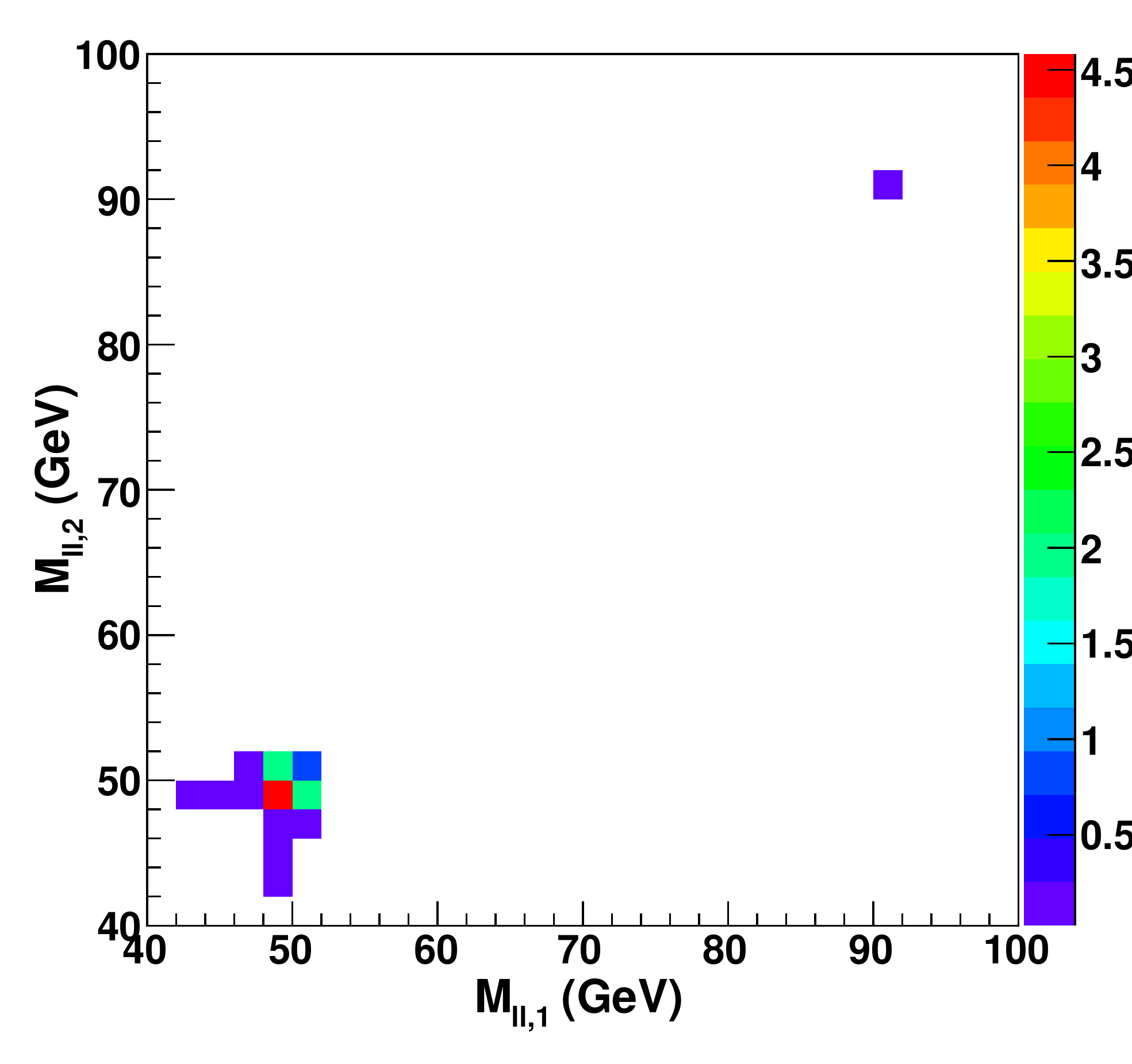}
\hspace{0.5cm}
\includegraphics[height=2.75in]{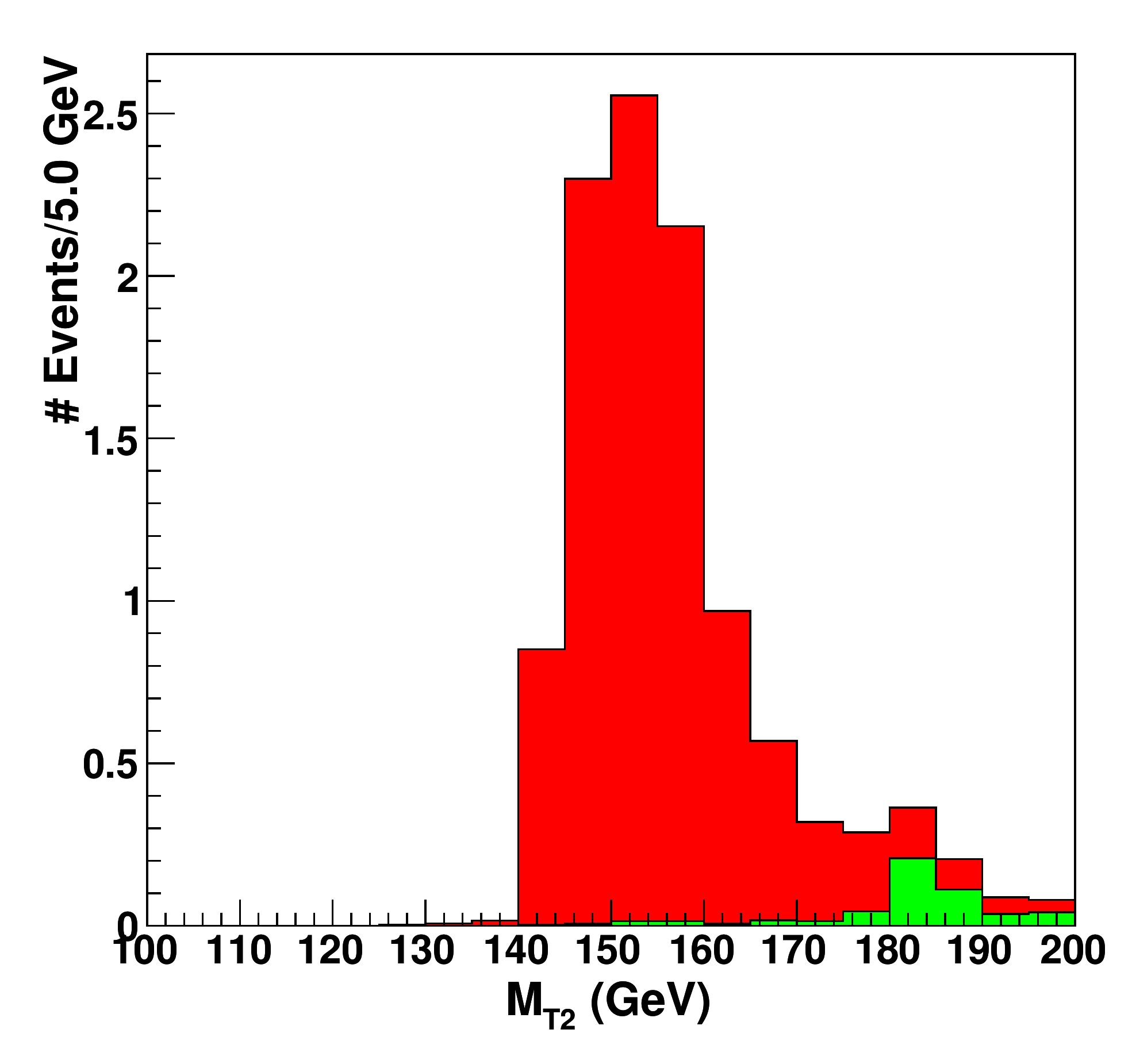}
\caption{$Z'$ signal (left panel) and the $M_{T_2}$ distribution (right panel) for the benchmark supersymmetry point. We have assumed a $7\tev$ collider and an integrated luminosity of $5\fb^{-1}$ (color online). The final state is $4\,\ell + \slashchar E_T$, which has no sizable standard model backgrounds. We have imposed a missing energy cut $\slashchar E_T > 20\gev$, which almost completely removes the $Z$ events from the plots. The number of events is smaller than in the Higgs case because we have to produce two neutralinos, rather than just one Higgs, to get the final state we want. The $M_{T_2}$ distribution was created using the true value of $M_{\tilde S}$. The distribution begins at $\sim M_{\tilde S} + M_{Z'}$ and has a drop-off at $\sim 160\gev \sim M_{\tilde N}$. The shape of $m_{T_2}$ -- something resembling a peak rather than a turn-on plus edge -- is an artifact of the small mass difference between $\sim M_{\tilde N}$ and $M_{\tilde S} + M_{Z'}$. The tail of the distribution comes from events where a $Z'$ was attributed to the incorrect parent neutralino. The small clump of events around $180\gev$ comes from $ZZ$ background which survived the $\ETmiss$ cut. }
\label{fig:susy_signs}
\end{figure*}

Requiring large missing energy, in addition to $4\,\ell$, gives us the ultimate clean signal. The SM $ZZ \rightarrow 4\,\ell$ events which constitute the only background in $4\,\ell$ channel do not survive a sufficiently large $\ETmiss$ cut. Such a clean environment is the ideal laboratory to study complicated signals, such as from supersymmetry. In the MSSM, it is hard to utilize this clean $4\,\ell + \ETmiss$ environment as only a small fraction of signal events have the required ingredients (there, however, exists a class of models with suitably designed mass hierarchies that result in multi-lepton and missing energy signals~\cite{DeSimone:2008gm,DeSimone:2009ws,Katz:2009qx,Katz:2010xg}). 
However, once the MSSM is extended by a Higgsphilic $Z'$, many avenues to $4\,\ell + \ETmiss$ open up instantly. In this section, we investigate the discovery potential of the bare-bones model in the $4\,\ell + \ETmiss$ channel, concentrating on the following topology 
\begin{equation}
  \label{eq:susy-topo}
  p p \text{ or } p \bar{p} \rightarrow \tilde N \tilde N  \rightarrow Z' Z' + \tilde S \tilde S \rightarrow 4\,\ell + \ETmiss
\end{equation}
As in previous subsections,  we first use the Tevatron data ($p \bar{p}$ initial state) to constrain the production cross-section of $4\,\ell + \ETmiss$, then discuss discovery potential at the LHC in the allowed region.  

There is no dedicated $4\,\ell\,+\,\slashchar E_T$ search at the Tevatron. The closest search is Ref.~\cite{CDF:10238}, the same $4\,\ell$ analysis we used to bound $h\rightarrow Z'Z'$ from Higges. While this search focused on exclusive $4\,\ell$ states, events with missing energy were not vetoed, so this search also limits $Z'$ produced from the decay of supersymmetric particles.
Using the same logic as in the $h\rightarrow Z'Z'$ section, we can bound the production cross section of certain supersymmetric particles. Assuming that no events have been found at the Tevatron with $5\fb^{-1}$ of data, we find that (as usual, at $95\%$CL)
\begin{equation}
  \label{eq:susy-bound}
  \begin{split}
    \sigma \left(p\bar p \rightarrow  \tilde N \tilde N \rightarrow 4\,\ell + \ETmiss \right)  \ & \leq 
                    \ 5.7\fb \left( \frac{ \mathcal{A}_{\text{sm}} } { \mathcal{A}_{\text{susy}} }\right)   \\
         & \simeq   4.3\fb  \; . 
  \end{split}
\end{equation}
where the value $5.7\fb$ comes directly from Eq.~\eqref{eq:cdf-lim-br} once the $BR(Z'\rightarrow \ell\ell)$ factors are removed. The $\mathcal A_{\text{susy}}$ in Eq.~\eqref{eq:susy-bound} accounts for any difference in acceptance between SM $Z \rightarrow \ell\ell$ events and supersymmetric $Z'$ events.  We have evaluated $\mathcal{A}_{\text{sm}}  / \mathcal{A}_{\text{susy}} $ using a set of events generated with a MadGraph implementation of the bare-bones model.  We find the basic kinematic acceptance of supersymmetric $Z'$ events to be slightly higher ($23\%$) than for Higgs-induced $Z'$ while the $m_{4\,\ell}$ cut remains approximately equally efficient. 

Let us assume that  $ M_{Z} > ( M_{\tilde N}  - M_{\tilde S} ) > M_{Z'}$. This choice of parameters ensures that $\tilde N \rightarrow \tilde S + Z'$ is almost $100\%$. It is particularly useful because, in this case, bounds on $4\,\ell + \ETmiss$ can be directly converted to bounds on $\tilde N$ masses.  If the branching fraction of $\tilde N$ to $Z'$ is less than $100\%$, all our constraints are on production cross-section times the branching fraction.
\begin{equation}
\begin{split}
\label{eq:N-bound}
 \sigma \left( p\bar p \rightarrow   \tilde N \tilde N  \right) &\times \Big( \text{Br}\left( Z' \rightarrow 2\ell\right)\Big)^2  \:  \leq 4.3\fb  \\ 
\Rightarrow  \quad & \quad M_{\tilde N}  \gtrsim 151\gev. 
\end{split}
\end{equation}
If the splitting among $\tilde N$ and $\tilde C$ is small, as is often the case, Eq.~\eqref{eq:N-bound} also constrains $M_{\tilde C}$. 

Another Tevatron measurement that is often used to constrain the color-neutral sector of supersymmetric models is the bound on tri-lepton/di-lepton + track production~\cite{CDF:9817}. This limit is most constraining when the sleptons are light and charginos/(heavier) neutralinos decay to $\tilde{\ell} + \nu/\tilde{\ell} + \ell$, essentially $100\%$ of the time. In the bare-bones model,  $\tilde N-\tilde C$ decay to gauge bosons rather than to sleptons. Subsequent decays of gauge bosons do yield leptons sometime, but the branching fractions of $W/Z/Z'$ to leptons suppresses the $3\,\ell$ rate to well below the current limit as established in Eq.~\eqref{eq:N-bound}. 

Having established the Tevatron constraints on sparticle masses, we now show an example of bare-bones supersymmetry production at the LHC. The benchmark spectra we choose is
\begin{equation}
  \label{eq:bench}
   M_{Z'} = 55\gev \quad  M_{\tilde S} = 90\gev \quad  M_{\tilde N} = 160\gev \; .
\end{equation}
The results are shown in Fig.~\ref{fig:susy_signs}. This point was generated using the selection criteria we used for the Higgs-induced $Z'$ signal at the LHC in Sec.~\ref{sec:higgs_prompt} plus an additional missing energy requirement $\slashchar E_T > 20\gev$. While unnecessary for $Z'$ discovery, the $\slashchar E_T$ cut is useful for extracting information about the superpartners. As in the Higgs case (Fig.~\ref{fig:fourlep}), we show the $Z'$-pair candidates in the $M_{\ell\ell,1}$ vs. $M_{\ell\ell,2}$ plane.  However we show signal and  background events for  $5\fb^{-1}$  of  data at $7\tev$ rather than $1\fb^{-1}$.  More luminosity is necessary since the supersymmetric signal is smaller -- we need to produce two neutralinos to get two $Z'$ instead of producing just one Higgs. As each event contains two invisible particles, we cannot simply combine all leptons to reconstruct the parent Higgsinos. However, we can get some idea on the mass scale of $M_{\tilde N}$ by using transverse variables. The second plot of Fig.~\ref{fig:susy_signs} shows the generalized transverse mass variable $M_{T_2}$~\cite{Lester:1999tx, Barr:2003rg, Cheng:2008hk} for these events. The generalized transverse mass depends on the mass of the invisible particle ($M_{\tilde S}$ for the benchmark point), however for simplicity we assume $m_{\tilde S}$ is already known in our calculations. With this additional assumption, the endpoint of $M_{T_2}$ gives the mass of $\tilde N$.  If $M_{\tilde S}$ is not known, we could still get some insight into $M_{\tilde N}$ by scanning over a range of $M_{\tilde S}$ values and looking for kink features in the $\text{max}(M_{T_2})-M_{\tilde{S}}$ plane~\cite{Barr:2003rg}. More involved techniques will certainly require larger datasets. However, because we are using $M_{T_2}$ in a purely leptonic environment, this endpoint and/or kink features are less susceptible to initial state radiation or experimental smearing than in hadronic applications. 

\subsection{Signatures of long lived particle}
\label{sec:long-lived}

Having surveyed the signals for prompt $Z'$ from Higgs and supersymmetric decays, in this section we address how these signals are modified when the $Z'$ is long-lived on collider scales. Before we begin, two comments are in order. First, the displaced bursts of energy left by a long-lived $Z'$ are highly exotic and free of any background; discoverability is limited solely by how often and how reliably such bursts can be triggered. Therefore, for this section we are not restricted to the leptonic decays of $Z'$. Second, exploring prospects of discovery of Higgs via long-lived $Z'$  is {\em highly} detector dependent and a full simulation is necessary to extract a completely reliable number. Such elaborate work is beyond the scope of this work. However, an approximate estimation of the number of triggerable events in the ATLAS is feasible based on the report Ref.~\cite{Aad:1175196}. \\

 \begin{figure}[!ht]
\includegraphics[width=0.45\textwidth]{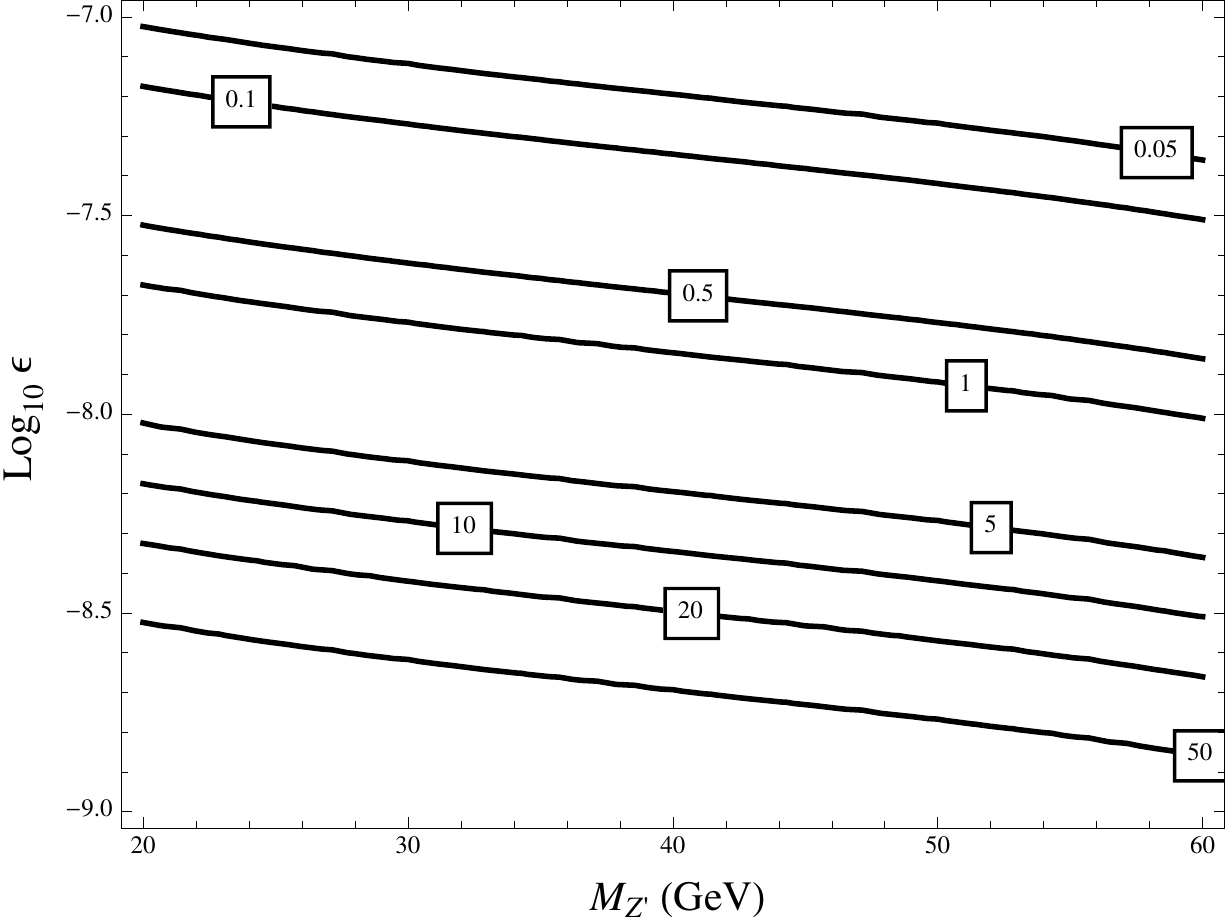}
\caption{ Contours showing the $Z'$ lifetime (in meters) as a function of the mixing parameter $\epsilon$ and the mass of $Z'$. }
\label{fig:life-time}
\end{figure}
The total decay width of $Z'$, and, therefore, the lifetime $\tau_{Z'}$ is sensitive on the value of $\epsilon$.  As long as $\epsilon \gtrsim 10^{-7}$,  its decay is mostly prompt and the conventional collider studies in previous sections apply. For a smaller $\epsilon$,  $Z'$ can be considered as long-lived and will decay at some macroscopic, displaced distance from the primary interaction. Since we are interested in decays within the detector volume, we are restricted to $\epsilon \gtrsim  10^{-9}$, corresponding to a decay length of $c \tau_{Z'} \lesssim 20~\text{m}$. This number comes from Ref.~\cite{Aad:1175196} and was based on the requirement that a sizable fraction of $Z'$ decays are subject to various triggers designed to find hidden valley particles~\footnote{Specifically, at $c \tau_{Z'} \lesssim 20~\text{m}$, at least $20\%$ of $Z'$ decays occur before the first Muon Spectrometer trigger plane in the ATLAS detector.}. For  even smaller mixing (or longer decay lengths), $Z'$s mostly decay outside the detector. Occasionally, a $Z'$ with $c\tau_{Z'} \gg 20\,m$ will decay inside the detector, though any realistic chance of seeing these events involves analyzing a much bigger sample of data than the first few inverse femtobarn samples.    

In Ref.~\cite{Aad:1175196}, a Higgs which decays mostly to $4\,b$ via a long-lived hidden sector particle was studied, and the article summarizes the performance and efficiency of three triggers: $(i)$ the Muon RoI cluster trigger $(ii)$ the $\log_{10}(E_\text{had}/E_\text{em})$ trigger, and $(iii)$ the ID-Trackless-jet+Muon trigger. The trigger are each designed to find sudden bursts of activity in different areas of the detector: The Muon RoI trigger looks for activity near the end of the HCal or in the Muon Spectrometer,  the $\log_{10}(E_\text{had}/E_\text{em})$ trigger looks at jets from decays inside the calorimeters, and the ID-Trackless-jet+Muon trigger looks for jets stemming from the Inner Detector beyond the first pixel layer and contains muons in the cone ($b$ or $c$ jets). 

\begin{figure}[!ht]
\centering
\includegraphics[width=3.0 in]{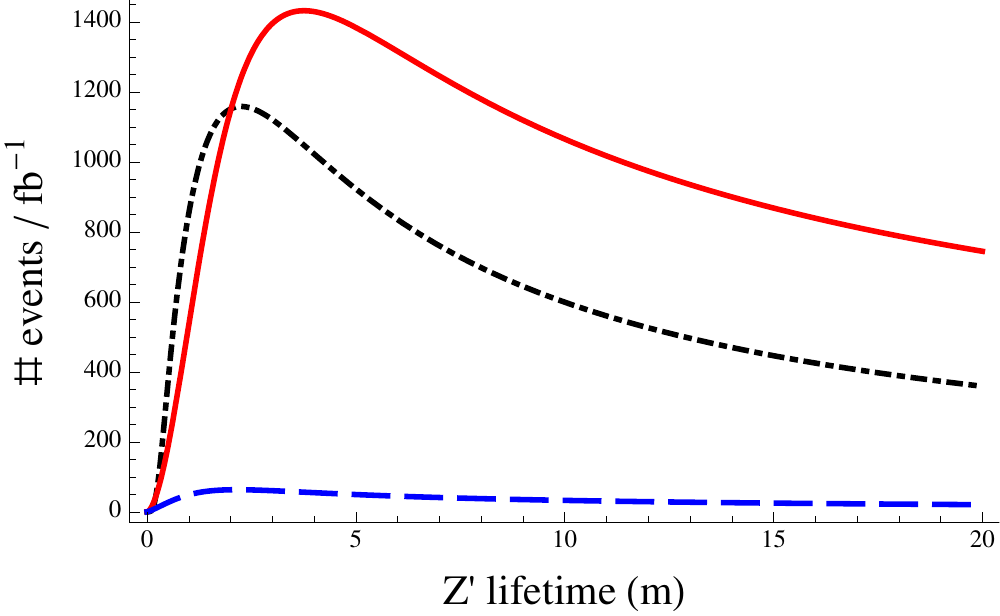}
\caption{ Number of $h \rightarrow Z'Z' \rightarrow \text{hadrons}$ events to be captured by the existing long-lived triggers  in the ATLAS detector. The red(solid), black(dot-dashed), and blue(dashed) lines refer to events captured by the $\log_{10}(E_\text{had}/E_\text{em})$ trigger, the ``muon RoI cluster trigger'' and the ``ID-trackless-jet + muon trigger'' respectively. We have assumed the Higgs mass to be $120\gev$ and BR$(h \rightarrow Z'Z')$ to be $100\%$. }
\label{fig:longlive}
\end{figure}
To estimate the number of triggerable events from the study done in Ref.~\cite{Aad:1175196} we concentrate only on the hadronic decays of $Z'$ as well as on the $Z'$ decays to hadronic $\tau$s. We also assume that $c-$hadrons give rise to muons at the same rate as the $b-$hadrons. In Fig.~\ref{fig:longlive}, we have plotted the number of triggerable events by various trigger objects at the ATLAS detector for $m_h = 120\gev$ and BR$(h \rightarrow Z'Z')$ to be $100\%$ at the LHC running  at $\sqrt{s} = 7\tev$.  The ID-trackless-jet + muon trigger is the least efficient trigger since  $c$ or $b$ hadrons decay to muons only $15-20\%$, even though $Z'$ decays to  $c$ or $b$ $\sim 27\%$.
  
We do not give a similar estimation $Z'$ decaying to leptons because of the lack of a detailed study as performed in Ref.~\cite{Aad:1175196}.  Assuming a flat trigger rate of $20\%$, we find that around $1300$ events per inverse$\fb$ will be triggered. However, a dedicated simulation is definitely warranted in order to estimate any realistic number. 

\section{From Viable Gravity Mediation to the Bare-bones Model}
\label{sec:viable-grav-medi}

Having surveyed how supersymmetric and Higgs phenomenology are altered by a Higgsphilic $Z'$, in this section we connect the bare-bones model to a particular limit of the viable gravity-mediated model of Ref.~\cite{Kribs:2010md}. 

\subsection{Viable Gravity Mediation in a Nutshell }
\label{sec:spectr-viable-grav}

Firstly, let us point out the key features in the spectrum of the viable mediation that distinguishes it from a typical MSSM study point.  
\begin{itemize}
\item  Gauginos are Dirac: they acquire masses along with fermionic components of superfields in the adjoint representation of the corresponding gauge group
~\cite{Polchinski:1982an,Hall:1990hq,Randall:1992cq,Fox:2002bu,Nelson:2002ca,Chacko:2004mi,Carone:2005iq,Nomura:2005rj,Nomura:2005qg,Carpenter:2005tz,Antoniadis:2006uj,Nakayama:2007cf,Amigo:2008rc}.  
\item  Soft squared-masses of the scalar superpartners of the SM fermions are {\emph not} flavor diagonal. There are neither $a-$type terms nor $\mu-$type Higgsino mass terms and, as a consequence, there are no left-right mixings among scalar soft-masses~\cite{Kribs:2007ac,Kribs:2009zy,Fok:2010vk}.     
\item  Gauginos are typically heavier than all other scalars.  
\item  The extended Higgs sector is charged under the $\SUL \times \UY \times \UX$ gauge groups which are spontaneously broken down to the $\Uem$ at the electroweak scale. The resultant spectrum contains the usual $W_\pm, Z$ and an additional massive charge neutral vector boson $Z'$.    
\end{itemize}

In the remaining part of this subsection we discuss the Higgs sector in detail as we show later that the effective description of it gradually becomes the bare-bones model constructed in Sec.~\ref{sec:model}.  Below we have listed all the particles in the Higgs sector along with their charges: 
\begin{equation}
\begin{tabular}{ c|c|c|c|c }
                   & $\SUC$    & $\SUL$    & $\UY$     & $\UX$    \\
\hline
$H_{u,d}$          & ${\bf 1}$ & ${\bf 2}$ & $\pm 1/2$ & $0$    \\
$R_{u,d}$          & ${\bf 1}$ & ${\bf 2}$ & $\mp 1/2$ & $\pm 1$  \\ 
$S_{u,d}$          & ${\bf 1}$ & ${\bf 1}$ & $0$       & $\mp 1$  \\ 
$T_{u,d}$          & ${\bf 1}$ & ${\bf 1}$ & $0$       & $\pm 2$  \\ 
\hline
\end{tabular} \; .
\label{H-charges}
\end{equation}
We use capitalized letters to designate the complete chiral supermultiplets,  small letters for the scalar component,  and capital letters with tildes for the fermionic  components of the same supermultiplet. 

The superpotential in the model consistent with the charges listed in eq.~\eqref{H-charges} 
\begin{equation}
  \begin{split}
  W \ \supset \ 
      \alpha_u \: S_u R_u H_u \ 
                   + \ \alpha_d \: S_d R_d H_d \ +  \\ 
     \qquad \qquad \qquad \frac{1}{2} \beta_u \: T_u S_u^2 
            \  + \ \frac{1}{2} \beta_d \: T_d S_d^2    \; ,
   \end{split}
   \label{eq:marginal-op}
\end{equation}
where we have neglected the usual Yukawa terms. The scalar potential contains the usual squared-soft masses for all multiplets as well as new $b-$type soft masses because of additional vector-like chiral supermultiplets. 
\begin{equation}
  \label{eq:int-higgs}
  \begin{split}
    V_{\text{soft} } \ \subset \ \sum_{\phi} \left( m_{\phi_u}^2 \: \phi_u^* \phi_u  \ + \ 
                  m_{\phi_d}^2 \: \phi_d^* \phi_d \right) \\ 
                            - \ \sum_{\phi} \left(   b_{\phi} \:  \phi_u \phi_d \ + \ \text{c.c.} \right)   \; ,
  \end{split}
\end{equation}
where field $\phi_{u,d}$ runs over the set of fields $ \{ h_{u,d}, r_{u,d}, s_{u,d}, s_{u,d}\}$: the scalar components of the corresponding  chiral supermultiplets. The Higgs potential  is minimized  around $\avg{h_{u,d}} \equiv v_{u,d} \neq 0$ and $\avg{s_{u,d}} \equiv v_{s_{u,d}} \neq 0$ and $\avg{r_{u,d}} = \avg{t_{u,d}} = 0$.  As a result 
the electroweak symmetry is broken down to electromagnetism ($\SUL \times \UY \rightarrow \Uem$). The additional symmetry  $\UX$ is broken completely which results in a massive $Z'$.  

\subsection{To the Bare-bones Model}
\label{sec:eff2bare-boness}

In order to extract the bare-bones model from the viable gravity mediation spectrum, let us consider the following set of limits:
\begin{equation}
  \label{eq:hscalar-limits}
\begin{split}
   g_X  \ \ll   \ g\, \text{ or } g'  \; ,  & \quad   M_{Z'} \ \lesssim M_{Z/W} \\
      \alpha_{u,d}  \ \simeq \   \beta_{u,d} \ &  \simeq \ \left(0.\text{few } - 1 \right) \\
  \tan \beta \equiv \frac{v_u}{v_d}  \gtrsim 3  \; , & \quad  \tan \beta_s \equiv \frac{v_{s_u}}{v_{s_d}} \sim 1  \; .
 \end{split}
\end{equation}
The requirement of large $\tan \beta$ limit is more general than just to derive the bare-bones model. It is necessary in order to  push the CP-even lightest Higgs boson above the LEP limit.

\subsubsection*{Gauge bosons, gauginos and matter superpartners}
\label{eff-gauge}
The $Z'$-mass is given by:
\begin{equation}
  \label{eq:MZ'} 
     M_{Z'} \ = \ \sqrt 2\, g_X  v_{s} \; , \quad  \text{where} \quad 
     v_s^2 \ = \ \left(   v_{s_u}^2 +  v_{s_d}^2 \right) \; .
\end{equation}
The implication of a small $g_X$  in Eq.~\eqref{eq:hscalar-limits} is that $M_{Z'} \ll v_s$.  In fact, since we seek $Z'$ at the electroweak scale, $v_s \gg M_Z, M_W$ and all degrees of freedom at or above $v_s$ can be easily integrated out. 

Also, note that at one-loop level, kinetic mixing of $\UY$ and $\UX$ is inevitably induced from the loops of $R_{u,d}$ which carry both charges. This generates the operator shown in Eq.~\eqref{kin-mix}. The mixing parameter $\epsilon$ in the viable gravity mediation is given by 
\begin{equation}
  \label{eq:epsilon}
  \epsilon \simeq 9 \times 10^{-4} \left(\frac{g_X}{0.1} \right) \log \frac{\Lambda}{1\tev} \;
\end{equation}
where $\Lambda$ denotes the scale above which there is no mixing (for example, the scale at which a non-Abelian group is broken to $\UX$). 
 
The Dirac nature of gauginos has a profound implication in the squared-masses of sleptons and squarks: the scalar masses do not get log-enhanced contributions due to the gaugino masses at one loop. As a result, the pattern of  superpartner masses is quite different from the typical MSSM spectra. Gluinos are the heaviest particles and are significantly heavier than the rest of the superpartners. Other gauginos have mass comparable to the squark and slepton masses, which, in turn, are of the order of scalar vev $v_s$. We therefore integrate out all gauginos, squarks and sleptons from the electroweak effective theory.

\subsubsection*{Scalar Higgs Multiplets}

Next, we turn our attention to the scalars in the Higgs sector. The scalars in the superfields $R_{u,d}$ or $T_{u,d}$ do not participate in the breaking of any symmetries and get physical masses simply from positive mass-squareds $m^2_{r_u, r_d}$ and  $m^2_{t_u, t_d}$ respectively. We can easily decouple them along with squarks and sleptons. 

Analysis of the neutral degrees of freedom in $H_{u,d}$ or $S_{u,d}$ is more involved, since all four states get non-zero vevs and mix among themselves.  The gauge-eigenstate fields can be expressed in terms of the mass eigenstate fields as:
\begin{equation}
  \label{eq:h-expn}
  \begin{split}
    \begin{pmatrix}
      h_u^0 \\ h_d^0 \\ s_u \\ s_d 
    \end{pmatrix}
   \ = \     \begin{pmatrix}
      v_u \\ v_d \\ v_{s_u} \\ v_{s_d} 
    \end{pmatrix} & 
       \  + \ \frac{1}{\sqrt 2}\ \mathcal{O}_1
   \begin{pmatrix}
      h  \\ H \\ s  \\ S 
    \end{pmatrix}  
\\     +  \  \frac{i}{\sqrt 2} \ &  \mathcal{O}_2 
   \begin{pmatrix}
      G_0 \\ A_h  \\ 0 \\ 0 
    \end{pmatrix} 
+ \   \frac{i}{\sqrt 2}\  \mathcal{O}_3
   \begin{pmatrix}
      0 \\ 0 \\ G'_0 \\ A_s   
    \end{pmatrix} \; .
  \end{split}
\end{equation}
In the above, $ \mathcal{O}_{1,2,3}$ are orthogonal rotation matrices, $G_0$ and $G'_0$ are the would-be Nambu- Goldstone bosons, which become the longitudinal modes of the $Z$ and $Z'$ massive vector bosons respectively, $A_h, A_s$ are the CP-odd states and finally $h,H,s,S$ are the CP-even states. 

In the limits of Eq.~\eqref{eq:hscalar-limits}, one also finds that the masses of the pseudo-scalars ($A_h, A_s$) are greater than or in the order of $v_s$, and there is minimal mixing between $h-$type states and $s-$type states. The mixing matrix can very well be approximated to be:
\begin{equation}
  \label{eq:defn-o}
  \mathcal{O}_1 \sim 
    \begin{pmatrix}
      \cos \alpha  & \sin \alpha & X & X \\ 
      -\sin \alpha & \cos \alpha & X & X  \\
       X & X &  \cos \alpha_s  & \sin \alpha_s  \\ 
       X & X & -\sin \alpha_s & \cos \alpha_s 
    \end{pmatrix}  \; ,
\end{equation}
where by $X$ we represent small parameters, the squares of which can be disregarded.  As a result of Eq.~\eqref{eq:hscalar-limits} the spectrum of CP-even scalars contains only one field ($h$) with mass lighter than $M_Z$ at the tree level.  In the limit that $v_s$ is large while the $\alpha$ and $\beta$ parameters in Eq.~\eqref{eq:hscalar-limits} are $\mathcal{O}(0.\text{few})$, both the scalars $s \text{ and }S$ are heavy. Similarly,  a large mass of the pseudo-scalar $A_h$ implies a large mass for $H$. 

One implication of the large $A_h$ mass and the specific form of the rotation matrix $\mathcal{O}_1 $ in Eq.~\eqref{eq:defn-o} is that the lightest CP-even Higgs scalar is similar to the MSSM Higgs in the decoupling limit - all its coupling to the SM fermions and the electroweak  gauge bosons are SM like to the leading order.  We additionally obtain the operator in Eq.~\eqref{int-hz'z'}, because of the non-vanishing $3\!\!-\!\!1$ and $4\!\!-\!\!1$ elements of $\mathcal{O}_1$ (call them $\delta_1$ and $\delta_2$ respectively) with the coefficient:
\begin{equation}
  \label{eq:g0}
  g_0 \ =\ g_X ( \delta_1 \sin \left( \beta_s \right)  + \delta_2 \cos \left( \beta_s\right))
\end{equation}

Turning to the charged Higgses, in the limit  of Eq.~\eqref{eq:hscalar-limits}, they generically have masses similar to that of the pseudo scalar $A_h$ and, hence, decouple. In case of large mixing one of the charged scalars might become light in the order of the EWSB scale. However, for the kind of phenomenology we discuss, this is mostly irrelevant. 

\subsubsection*{Charged and Neutral Higgsinos}
Lastly, we come to the charginos and neutralinos. Expanding the superpotential in Eq.~\eqref{eq:marginal-op} around the Higgs vevs, we find the chargino masses to be: 
\begin{equation}
  \label{eq:chr-mass}
        \alpha_u \: v_s \sin \beta_X   \: R_u^{-}  \tilde H_u^{+} \ 
                   + \ \alpha_d \:  v_s \cos \beta_X  \:  H_d^{-} R_d^{+} \; .
\end{equation}
Note that if either of the product $\alpha_u \sin \beta_X$ or $\alpha_d \cos \beta_X$ is bigger than the other, one of the chargino decouples.  Without loss of generality we assume that 
 $\alpha_u \:  \sin \beta_X  \ll \alpha_d  \cos \beta_X$ and we integrate out the charginos 
$H_d^{-}-R_d^{+}$.  The Dirac chargino $\tilde C$ in our bare-bones model is, therefore, represented by  
\begin{equation}
  \label{eq:defn-C}
  \tilde C \ = \
   \begin{pmatrix}
     \tilde H_u^+ \\ \tilde R_{u}^{- *} 
   \end{pmatrix} \; . 
\end{equation}

The neutral Higgsino sector is slightly more complicated, since we need to deal with eight (two component) neutralinos. However, because of the pure Dirac nature of the neutralinos, the mass matrix breaks up into $4\times 4$ blocks. Each $4\times 4$ neutralino mass matrix can be further reduced to $2\times 2$ blocks since the $U-$type and $D-$type spinors are totally disconnected%
~\footnote{ The  $ U\leftrightarrow D$ symmetry in the superpotential shown in Eq.~\eqref{eq:marginal-op} results in $U$ and $D$ sector being completely disjoint to the leading order.  This is however not true at higher orders:   
because of Higgs vevs $v_{u,d}$, a tiny mixing between the $U$ and $D$ sector of order $g_i^2  v_{u}v_{d}/M_i^2$ gets generated, where $M_i$ refers to the gaugino mass of the $i-$th gauge group.}.
Before proceedings let us specify the full neutralino mass matrix:  
\begin{equation}
  \label{eq:neut-mass}
     \begin{bmatrix}
  \tilde H_u^0  & \tilde S_u^0 
  \end{bmatrix} \!\!
  \begin{bmatrix}
    \alpha_u \: v_s \sin \beta_X & 0  \\
     \alpha_u \: v  \sin \beta  &    \beta_u \: v_s \sin \beta_X   
  \end{bmatrix}  \!\!
 \begin{bmatrix}
  \tilde R_u^0  \\  \tilde T_u^0      
  \end{bmatrix}   +   u\rightarrow d
 \end{equation}
In the limit  $\alpha_u \:  \sin \beta_X  \ll \alpha_d  \cos \beta_X$ and  $\beta_u \:  \sin \beta_X  \ll \beta_d  \cos \beta_X$,  entire $d-$sector decouples and we need to worry about only four neutralinos $\tilde H_u, \tilde R_u, \tilde S_u, \tilde T_u$.  Also note that in the limit $v_s \gg v$, (implied by Eq.~\eqref{eq:hscalar-limits}) there is minimal mixing between the $\tilde H- \tilde R$ states and  $\tilde S - \tilde T$  states,  and the mixing can be treated as perturbation.  The Dirac neutralinos $\tilde N$
and $\tilde S$ in the bare-bones model are then simply given as 
\begin{equation}
  \label{eq:defn-NS}
    \tilde N \ = \
   \begin{pmatrix}
     \tilde H_u \\ \tilde R_{u}^{*} 
   \end{pmatrix} \text{ and }\quad 
    \tilde S \ = \
   \begin{pmatrix}
     \tilde S_u \\ \tilde T_{u}^{*} 
   \end{pmatrix}\; . 
\end{equation}

With the mapping established in Eqs.~\eqref{eq:defn-C} and \eqref{eq:defn-NS}, we can now derive the operators in Eq.~\eqref{int-nhz'} with the identification
\begin{equation}
  \label{eq:matching-NS}
  \begin{split}
    \eta \ = \  \left( \frac{\alpha_u^2}{\beta_u^2 - \alpha_u^2} \right)  \left( \frac{v \sin \beta}{v_s \sin \beta_X} \right) \\
   \zeta \ = \  \left( \frac{\alpha_u \beta_u}{\beta_u^2 - \alpha_u^2} \right)  \left( \frac{v \sin \beta}{v_s \sin \beta_X} \right) 
  \end{split}
\end{equation}

\subsection{A Numerical Example}
\label{sec:num}

In the previous subsections we have outlined the limits in which the bare-bones emerges as an effective theory of viable gravity mediation.  Following Ref.~\cite{Kribs:2010md}, it is easy to check that the gauginos and the matter superpartners are heavier than the electroweak scale, and needs no additional supporting argument. It is slightly nontrivial to analyze  neutral scalar Higgses (CP-even scalars $h,H,s,S$,  CP-odd scalars $A_h, A_s$), neutral and charged Higgsinos,  and then to recover the bare-bones model from it. To demonstrate the mapping discussed in Subsec.~\ref{sec:eff2bare-boness},  we find a point in the parameter space of the viable gravity mediation and show numerically that it reduces to  the benchmark point in Eq.~\eqref{eq:bench}. 

First of all, note that in order to fully specify the Higgs sector, we need four fewer parameters because of the four equalities that result from the minimization conditions $\partial V/ \partial v_{u,d}^0 = \partial V/ \partial v_{s_{u,d}} = 0$.   We use them to eliminate four input parameters, $m_{H_{u,d}}^2$, $m_{S_{u,d}}^2$ in terms of other soft and supersymmetry breaking parameters. 
Below we list the initial set of parameters that we use to  calculate the spectrum:
\begin{equation}
  \label{eq:param-in}
 \begin{split}
 &  \qquad v_s \: = \:   700\gev \>,\qquad   g_X \: = \: 0.06  \, ,  \\
 &    \qquad    \tan\beta = 1  \>,\qquad    \tan\beta_X = 10 \, , \\
 &  B_{h} =  (300\gev)^2 \>,\quad  B_s =  (300\gev)^2 \,,  \\
 &     \alpha_u = 0.3 \>,\quad \alpha_d = 0.6 \>,\quad 
                        \beta_u = 0.2\>,  \quad \beta_d = 0.7 \, .
\end{split}
\end{equation}

Let us first start with  the neutral scalar spectrum%
\footnote{Because of the Dirac nature of  gauginos, the D-term contributions to the scalar quartic are usually suppressed. In this work, however, we are assuming large squared soft masses for the scalar adjoints so that the suppression is negligible and we recover MSSM-like quartic.}: 
\begin{equation}
\begin{split}
  \label{eq:bench-scalar-mass}
& \qquad \quad m_h =  61\gev \>,\qquad   m_H =  954\gev \, , \\ 
& \qquad \quad  m_s =  231\gev \>, \qquad  m_S = 512\gev \, , \\ 
& \quad \quad m_{A_h} = 953\gev   \>, \qquad m_{A_s}  = 424\gev \, . 
  \end{split}
\end{equation}
The lightest CP-even Higgs $h$ obtain a tree level mass of only $61\gev$. Large contributions to its mass arise due to radiative corrections from top-stop loop and neutral Higgs/Higgsino loop, which help it to avoid the LEP  limit~\cite{Kribs:2010md}. 

Next, consider the Higgsino spectrum:
\begin{equation}
\begin{split}
  \label{eq:bench-ino-mass}
& \qquad \quad M_{\tilde S_u - \tilde T_u} =   91\gev \>,\qquad   M_{\tilde H_u - \tilde R_u} =  162\gev \, , \\ 
& \qquad \quad M_{\tilde S_d - \tilde T_d} =  347\gev \>,\qquad   M_{\tilde H_d - \tilde R_d} =  296\gev \, , \\ 
& \quad \quad  M_{\tilde H_u^+ - \tilde R_u^-} =  148\gev \>,\qquad  
                      M_{\tilde H_d^- - \tilde R_u^+} =  297\gev \, , \, . 
  \end{split}
\end{equation}
Considering only the particles with mass less than $200\gev$, we recover the benchmark point  in Eq.~\eqref{eq:bench} with the identifications as listed in Eqs.~\eqref{eq:defn-C} and \eqref{eq:defn-NS}.

\section{Conclusion}
\label{sec:conclusion}
The narrow width of a light Higgs boson ( $m_h \lesssim 130\gev$ ),  makes its decays particularly sensitive to new physics. Beyond-the-SM particles which interact with the Higgs, can open up important and even dominant Higgs decay modes while still coupling quite weakly. This susceptibility to new decay modes is particularly prevalent in weak-scale supersymmetry, where the underlying supersymmetric structure forces it to be light.  

The particular scenario we explore here is an extension of the minimal supersymmetric setup by a new and massive gauge boson $Z'$.  This $Z'$ has infinitesimal couplings to all matter, with the exception of  Higgses and Higgsinos. Higgses in this scenario can decay to two $Z'$ gauge bosons, each of which  subsequently decays to two leptons. Thus, with this small modification, we resurrect the gold-plated $h \rightarrow 4\,\ell$ decay mode for light Higgses. For maximum effect,  $Z'$ should neither be extremely light nor extremely heavy, a case which has been under-represented in $Z'$ phenomenology. 

A $Z'$ with exactly the right properties emerges automatically from the recent, viable gravity-mediated scenario of Ref.~\cite{Kribs:2010md}. Inspired by this model, we study Higgsphilic $Z'$ coming from the decays of supersymmetric particles, as well as $Z'$ coming from Higgs decay. For simplicity, and in the spirit of Ref.~\cite{Alwall:2008ag}, we perform these studies using a bare-bones version of the scenario in Ref~\cite{Kribs:2010md}, which contains only the phenomenologically important part of the Lagrangian.

For prompt $Z'$ decays we impose bounds from recent CDF $4\,\ell$ searches~\cite{CDF:10238}.  Percent level $h \rightarrow Z'Z'$ fractions are completely consistent with all bounds and can lead to impressive and unmistakable signals, including $5~\sigma$ Higgs and $Z'$ discovery within the first $\fb^{-1}$ of LHC.  We also discuss the possibilities for long-lived $Z'$ detection with $1\fb^{-1}$ of data. 

Within this setup, we also find $4\ell + \ETmiss$ to be a superb channel for supersymmetry discovery. As a conservative estimate, we study electroweak pair-production of neutral Higgsinos that emit $Z'$  as they decay to the LSPs,  leading to final states containing $Z'Z' + \ETmiss$.  As the Higgsinos need to be pair produced, Higgsino discovery in $4\ell + \ETmiss$ requires more luminosity than the Higgs discovery: $\gtrsim 5\fb^{-1}$ of integrated luminosity. However, given that the majority of all supersymmetric events will eventually cascade-decay into neutralinos, the total rate of supersymmetric events yielding $4\ell + \ETmiss$ is potentially much higher than the rate we considered here.

\section*{Acknowledgments}
We thank Graham Kribs for comments on the manuscript. TSR acknowledges private communications to Henry Lubatti. AM is supported by Fermilab operated by Fermi Research Alliance,  LLC under contract number DE-AC02-07CH11359 with the US Department of Energy. TSR is supported in part by the US Department of Energy  under contract numbers DE-FGO2-96ER40956. 

%\appendix*

\bibliography{references}

%merlin.mbs 2010-03-15 4.21a (PWD, AO, DPC)
\begin{thebibliography}{71}%
\makeatletter
\providecommand \@ifxundefined [1]{%
 \@ifx{#1\undefined}
}%
\providecommand \@ifnum [1]{%
 \ifnum #1\expandafter \@firstoftwo
 \else \expandafter \@secondoftwo
 \fi
}%
\providecommand \@ifx [1]{%
 \ifx #1\expandafter \@firstoftwo
 \else \expandafter \@secondoftwo
 \fi
}%
\providecommand \natexlab [1]{#1}%
\providecommand \enquote  [1]{``#1''}%
\providecommand \bibnamefont  [1]{#1}%
\providecommand \bibfnamefont [1]{#1}%
\providecommand \citenamefont [1]{#1}%
\providecommand \href@noop [0]{\@secondoftwo}%
\providecommand \href [0]{\begingroup \@sanitize@url \@href}%
\providecommand \@href[1]{\@@startlink{#1}\@@href}%
\providecommand \@@href[1]{\endgroup#1\@@endlink}%
\providecommand \@sanitize@url [0]{\catcode `\\12\catcode `\$12\catcode
  `\&12\catcode `\#12\catcode `\^12\catcode `\_12\catcode `\%12\relax}%
\providecommand \@@startlink[1]{}%
\providecommand \@@endlink[0]{}%
\providecommand \url  [0]{\begingroup\@sanitize@url \@url }%
\providecommand \@url [1]{\endgroup\@href {#1}{\urlprefix }}%
\providecommand \urlprefix  [0]{URL }%
\providecommand \Eprint [0]{\href }%
\@ifxundefined \urlstyle {%
  \providecommand \doi  [0]{\begingroup \@sanitize@url \@doi}%
  \providecommand \@doi [1]{\endgroup \@@startlink {\doibase
  #1}doi:\discretionary {}{}{}#1\@@endlink }%
}{%
  \providecommand \doi  [0]{doi:\discretionary{}{}{}\begingroup
  \urlstyle{rm}\Url }%
}%
\providecommand \doibase [0]{http://dx.doi.org/}%
\providecommand \Doi [0]{\begingroup \@sanitize@url \@Doi }%
\providecommand \@Doi  [1]{\endgroup\@@startlink{\doibase#1}\@@Doi}%
\providecommand \@@Doi [1]{#1\@@endlink}%
\providecommand \selectlanguage [0]{\@gobble}%
\providecommand \bibinfo  [0]{\@secondoftwo}%
\providecommand \bibfield  [0]{\@secondoftwo}%
\providecommand \translation [1]{[#1]}%
\providecommand \BibitemOpen [0]{}%
\providecommand \bibitemStop [0]{}%
\providecommand \bibitemNoStop [0]{.\EOS\space}%
\providecommand \EOS [0]{\spacefactor3000\relax}%
\providecommand \BibitemShut  [1]{\csname bibitem#1\endcsname}%
%</preamble>
\bibitem [{\citenamefont {Aad}\ \emph {et~al.}(2009){\natexlab{a}}\citenamefont
  {Aad} \emph {et~al.}}]{Aad:2009wy}%
  \BibitemOpen
  \bibfield  {author} {\bibinfo {author} {\bibfnamefont {G.}~\bibnamefont
  {Aad}} \emph {et~al.} (\bibinfo {collaboration} {The ATLAS Collaboration}),\
  }\href@noop {} { (\bibinfo {year} {2009}{\natexlab{a}})},\ \Eprint
  {http://arxiv.org/abs/0901.0512} {arXiv:0901.0512 [hep-ex]} \BibitemShut
  {NoStop}%
\bibitem [{\citenamefont {Carena}\ \emph {et~al.}(2000)\citenamefont {Carena},
  \citenamefont {Haber}, \citenamefont {Heinemeyer}, \citenamefont {Hollik},
  \citenamefont {Wagner} \emph {et~al.}}]{Carena:2000dp}%
  \BibitemOpen
  \bibfield  {author} {\bibinfo {author} {\bibfnamefont {M.~S.}\ \bibnamefont
  {Carena}}, \bibinfo {author} {\bibfnamefont {H.}~\bibnamefont {Haber}},
  \bibinfo {author} {\bibfnamefont {S.}~\bibnamefont {Heinemeyer}}, \bibinfo
  {author} {\bibfnamefont {W.}~\bibnamefont {Hollik}}, \bibinfo {author}
  {\bibfnamefont {C.}~\bibnamefont {Wagner}},  \emph {et~al.},\ }\Doi
  {10.1016/S0550-3213(00)00212-1} {\bibfield  {journal} {\bibinfo  {journal}
  {Nucl.Phys.},\ }\textbf {\bibinfo {volume} {B580}},\ \bibinfo {pages} {29}
  (\bibinfo {year} {2000})},\ \Eprint {http://arxiv.org/abs/hep-ph/0001002}
  {arXiv:hep-ph/0001002 [hep-ph]} \BibitemShut {NoStop}%
\bibitem [{\citenamefont {Kribs}\ \emph
  {et~al.}(2010){\natexlab{a}}\citenamefont {Kribs}, \citenamefont {Martin},
  \citenamefont {Roy},\ and\ \citenamefont {Spannowsky}}]{Kribs:2009yh}%
  \BibitemOpen
  \bibfield  {author} {\bibinfo {author} {\bibfnamefont {G.~D.}\ \bibnamefont
  {Kribs}}, \bibinfo {author} {\bibfnamefont {A.}~\bibnamefont {Martin}},
  \bibinfo {author} {\bibfnamefont {T.~S.}\ \bibnamefont {Roy}}, \ and\
  \bibinfo {author} {\bibfnamefont {M.}~\bibnamefont {Spannowsky}},\ }\Doi
  {10.1103/PhysRevD.81.111501} {\bibfield  {journal} {\bibinfo  {journal}
  {Phys.Rev.},\ }\textbf {\bibinfo {volume} {D81}},\ \bibinfo {pages} {111501}
  (\bibinfo {year} {2010}{\natexlab{a}})},\ \Eprint
  {http://arxiv.org/abs/0912.4731} {arXiv:0912.4731 [hep-ph]} \BibitemShut
  {NoStop}%
\bibitem [{\citenamefont {Kribs}\ \emph
  {et~al.}(2010){\natexlab{b}}\citenamefont {Kribs}, \citenamefont {Martin},
  \citenamefont {Roy},\ and\ \citenamefont {Spannowsky}}]{Kribs:2010hp}%
  \BibitemOpen
  \bibfield  {author} {\bibinfo {author} {\bibfnamefont {G.~D.}\ \bibnamefont
  {Kribs}}, \bibinfo {author} {\bibfnamefont {A.}~\bibnamefont {Martin}},
  \bibinfo {author} {\bibfnamefont {T.~S.}\ \bibnamefont {Roy}}, \ and\
  \bibinfo {author} {\bibfnamefont {M.}~\bibnamefont {Spannowsky}},\ }\Doi
  {10.1103/PhysRevD.82.095012} {\bibfield  {journal} {\bibinfo  {journal}
  {Phys.Rev.},\ }\textbf {\bibinfo {volume} {D82}},\ \bibinfo {pages} {095012}
  (\bibinfo {year} {2010}{\natexlab{b}})},\ \Eprint
  {http://arxiv.org/abs/1006.1656} {arXiv:1006.1656 [hep-ph]} \BibitemShut
  {NoStop}%
\bibitem [{\citenamefont {Kribs}\ \emph
  {et~al.}(2010){\natexlab{c}}\citenamefont {Kribs}, \citenamefont {Martin},\
  and\ \citenamefont {Roy}}]{Kribs:2010ii}%
  \BibitemOpen
  \bibfield  {author} {\bibinfo {author} {\bibfnamefont {G.~D.}\ \bibnamefont
  {Kribs}}, \bibinfo {author} {\bibfnamefont {A.}~\bibnamefont {Martin}}, \
  and\ \bibinfo {author} {\bibfnamefont {T.~S.}\ \bibnamefont {Roy}},\
  }\href@noop {} { (\bibinfo {year} {2010}{\natexlab{c}})},\ \Eprint
  {http://arxiv.org/abs/1012.2866} {arXiv:1012.2866 [hep-ph]} \BibitemShut
  {NoStop}%
\bibitem [{\citenamefont {Chang}\ \emph {et~al.}(2006)\citenamefont {Chang},
  \citenamefont {Fox},\ and\ \citenamefont {Weiner}}]{Chang:2005ht}%
  \BibitemOpen
  \bibfield  {author} {\bibinfo {author} {\bibfnamefont {S.}~\bibnamefont
  {Chang}}, \bibinfo {author} {\bibfnamefont {P.~J.}\ \bibnamefont {Fox}}, \
  and\ \bibinfo {author} {\bibfnamefont {N.}~\bibnamefont {Weiner}},\ }\Doi
  {10.1088/1126-6708/2006/08/068} {\bibfield  {journal} {\bibinfo  {journal}
  {JHEP},\ }\textbf {\bibinfo {volume} {08}},\ \bibinfo {pages} {068} (\bibinfo
  {year} {2006})},\ \Eprint {http://arxiv.org/abs/hep-ph/0511250}
  {arXiv:hep-ph/0511250} \BibitemShut {NoStop}%
%%CITATION = HEP-PH/0511250;%%
\bibitem [{\citenamefont {Carena}\ \emph {et~al.}(2008)\citenamefont {Carena},
  \citenamefont {Han}, \citenamefont {Huang},\ and\ \citenamefont
  {Wagner}}]{Carena:2007jk}%
  \BibitemOpen
  \bibfield  {author} {\bibinfo {author} {\bibfnamefont {M.}~\bibnamefont
  {Carena}}, \bibinfo {author} {\bibfnamefont {T.}~\bibnamefont {Han}},
  \bibinfo {author} {\bibfnamefont {G.-Y.}\ \bibnamefont {Huang}}, \ and\
  \bibinfo {author} {\bibfnamefont {C.~E.}\ \bibnamefont {Wagner}},\ }\Doi
  {10.1088/1126-6708/2008/04/092} {\bibfield  {journal} {\bibinfo  {journal}
  {JHEP},\ }\textbf {\bibinfo {volume} {0804}},\ \bibinfo {pages} {092}
  (\bibinfo {year} {2008})},\ \Eprint {http://arxiv.org/abs/0712.2466}
  {arXiv:0712.2466 [hep-ph]} \BibitemShut {NoStop}%
\bibitem [{\citenamefont {Chang}\ \emph {et~al.}(2008)\citenamefont {Chang},
  \citenamefont {Dermisek}, \citenamefont {Gunion},\ and\ \citenamefont
  {Weiner}}]{Chang:2008cw}%
  \BibitemOpen
  \bibfield  {author} {\bibinfo {author} {\bibfnamefont {S.}~\bibnamefont
  {Chang}}, \bibinfo {author} {\bibfnamefont {R.}~\bibnamefont {Dermisek}},
  \bibinfo {author} {\bibfnamefont {J.~F.}\ \bibnamefont {Gunion}}, \ and\
  \bibinfo {author} {\bibfnamefont {N.}~\bibnamefont {Weiner}},\ }\Doi
  {10.1146/annurev.nucl.58.110707.171200} {\bibfield  {journal} {\bibinfo
  {journal} {Ann. Rev. Nucl. Part. Sci.},\ }\textbf {\bibinfo {volume} {58}},\
  \bibinfo {pages} {75} (\bibinfo {year} {2008})},\ \Eprint
  {http://arxiv.org/abs/0801.4554} {arXiv:0801.4554 [hep-ph]} \BibitemShut
  {NoStop}%
%%CITATION = 0801.4554;%%
\bibitem [{\citenamefont {Lisanti}\ and\ \citenamefont
  {Wacker}(2009)}]{Lisanti:2009uy}%
  \BibitemOpen
  \bibfield  {author} {\bibinfo {author} {\bibfnamefont {M.}~\bibnamefont
  {Lisanti}}\ and\ \bibinfo {author} {\bibfnamefont {J.~G.}\ \bibnamefont
  {Wacker}},\ }\Doi {10.1103/PhysRevD.79.115006} {\bibfield  {journal}
  {\bibinfo  {journal} {Phys. Rev.},\ }\textbf {\bibinfo {volume} {D79}},\
  \bibinfo {pages} {115006} (\bibinfo {year} {2009})},\ \Eprint
  {http://arxiv.org/abs/0903.1377} {arXiv:0903.1377 [hep-ph]} \BibitemShut
  {NoStop}%
%%CITATION = 0903.1377;%%
\bibitem [{\citenamefont {Arkani-Hamed}\ \emph {et~al.}(2009)\citenamefont
  {Arkani-Hamed}, \citenamefont {Finkbeiner}, \citenamefont {Slatyer},\ and\
  \citenamefont {Weiner}}]{ArkaniHamed:2008qn}%
  \BibitemOpen
  \bibfield  {author} {\bibinfo {author} {\bibfnamefont {N.}~\bibnamefont
  {Arkani-Hamed}}, \bibinfo {author} {\bibfnamefont {D.~P.}\ \bibnamefont
  {Finkbeiner}}, \bibinfo {author} {\bibfnamefont {T.~R.}\ \bibnamefont
  {Slatyer}}, \ and\ \bibinfo {author} {\bibfnamefont {N.}~\bibnamefont
  {Weiner}},\ }\Doi {10.1103/PhysRevD.79.015014} {\bibfield  {journal}
  {\bibinfo  {journal} {Phys.Rev.},\ }\textbf {\bibinfo {volume} {D79}},\
  \bibinfo {pages} {015014} (\bibinfo {year} {2009})},\ \Eprint
  {http://arxiv.org/abs/0810.0713} {arXiv:0810.0713 [hep-ph]} \BibitemShut
  {NoStop}%
\bibitem [{\citenamefont {Cheung}\ \emph {et~al.}(2009)\citenamefont {Cheung},
  \citenamefont {Ruderman}, \citenamefont {Wang},\ and\ \citenamefont
  {Yavin}}]{Cheung:2009qd}%
  \BibitemOpen
  \bibfield  {author} {\bibinfo {author} {\bibfnamefont {C.}~\bibnamefont
  {Cheung}}, \bibinfo {author} {\bibfnamefont {J.~T.}\ \bibnamefont
  {Ruderman}}, \bibinfo {author} {\bibfnamefont {L.-T.}\ \bibnamefont {Wang}},
  \ and\ \bibinfo {author} {\bibfnamefont {I.}~\bibnamefont {Yavin}},\ }\Doi
  {10.1103/PhysRevD.80.035008} {\bibfield  {journal} {\bibinfo  {journal}
  {Phys.Rev.},\ }\textbf {\bibinfo {volume} {D80}},\ \bibinfo {pages} {035008}
  (\bibinfo {year} {2009})},\ \Eprint {http://arxiv.org/abs/0902.3246}
  {arXiv:0902.3246 [hep-ph]} \BibitemShut {NoStop}%
\bibitem [{\citenamefont {Morrissey}\ \emph {et~al.}(2009)\citenamefont
  {Morrissey}, \citenamefont {Poland},\ and\ \citenamefont
  {Zurek}}]{Morrissey:2009ur}%
  \BibitemOpen
  \bibfield  {author} {\bibinfo {author} {\bibfnamefont {D.~E.}\ \bibnamefont
  {Morrissey}}, \bibinfo {author} {\bibfnamefont {D.}~\bibnamefont {Poland}}, \
  and\ \bibinfo {author} {\bibfnamefont {K.~M.}\ \bibnamefont {Zurek}},\ }\Doi
  {10.1088/1126-6708/2009/07/050} {\bibfield  {journal} {\bibinfo  {journal}
  {JHEP},\ }\textbf {\bibinfo {volume} {0907}},\ \bibinfo {pages} {050}
  (\bibinfo {year} {2009})},\ \Eprint {http://arxiv.org/abs/0904.2567}
  {arXiv:0904.2567 [hep-ph]} \BibitemShut {NoStop}%
\bibitem [{\citenamefont {Strassler}\ and\ \citenamefont
  {Zurek}(2007)}]{Strassler:2006im}%
  \BibitemOpen
  \bibfield  {author} {\bibinfo {author} {\bibfnamefont {M.~J.}\ \bibnamefont
  {Strassler}}\ and\ \bibinfo {author} {\bibfnamefont {K.~M.}\ \bibnamefont
  {Zurek}},\ }\Doi {10.1016/j.physletb.2007.06.055} {\bibfield  {journal}
  {\bibinfo  {journal} {Phys.Lett.},\ }\textbf {\bibinfo {volume} {B651}},\
  \bibinfo {pages} {374} (\bibinfo {year} {2007})},\ \Eprint
  {http://arxiv.org/abs/hep-ph/0604261} {arXiv:hep-ph/0604261 [hep-ph]}
  \BibitemShut {NoStop}%
\bibitem [{\citenamefont {Strassler}(2006)}]{Strassler:2006qa}%
  \BibitemOpen
  \bibfield  {author} {\bibinfo {author} {\bibfnamefont {M.~J.}\ \bibnamefont
  {Strassler}},\ }\href@noop {} { (\bibinfo {year} {2006})},\ \Eprint
  {http://arxiv.org/abs/hep-ph/0607160} {arXiv:hep-ph/0607160 [hep-ph]}
  \BibitemShut {NoStop}%
\bibitem [{\citenamefont {Han}\ \emph {et~al.}(2008)\citenamefont {Han},
  \citenamefont {Si}, \citenamefont {Zurek},\ and\ \citenamefont
  {Strassler}}]{Han:2007ae}%
  \BibitemOpen
  \bibfield  {author} {\bibinfo {author} {\bibfnamefont {T.}~\bibnamefont
  {Han}}, \bibinfo {author} {\bibfnamefont {Z.}~\bibnamefont {Si}}, \bibinfo
  {author} {\bibfnamefont {K.~M.}\ \bibnamefont {Zurek}}, \ and\ \bibinfo
  {author} {\bibfnamefont {M.~J.}\ \bibnamefont {Strassler}},\ }\Doi
  {10.1088/1126-6708/2008/07/008} {\bibfield  {journal} {\bibinfo  {journal}
  {JHEP},\ }\textbf {\bibinfo {volume} {0807}},\ \bibinfo {pages} {008}
  (\bibinfo {year} {2008})},\ \Eprint {http://arxiv.org/abs/0712.2041}
  {arXiv:0712.2041 [hep-ph]} \BibitemShut {NoStop}%
\bibitem [{\citenamefont {Falkowski}\ \emph {et~al.}(2010)\citenamefont
  {Falkowski}, \citenamefont {Ruderman}, \citenamefont {Volansky},\ and\
  \citenamefont {Zupan}}]{Falkowski:2010gv}%
  \BibitemOpen
  \bibfield  {author} {\bibinfo {author} {\bibfnamefont {A.}~\bibnamefont
  {Falkowski}}, \bibinfo {author} {\bibfnamefont {J.~T.}\ \bibnamefont
  {Ruderman}}, \bibinfo {author} {\bibfnamefont {T.}~\bibnamefont {Volansky}},
  \ and\ \bibinfo {author} {\bibfnamefont {J.}~\bibnamefont {Zupan}},\
  }\href@noop {} { (\bibinfo {year} {2010})},\ \Eprint
  {http://arxiv.org/abs/1007.3496} {arXiv:1007.3496 [hep-ph]} \BibitemShut
  {NoStop}%
%%CITATION = 1007.3496;%%
\bibitem [{\citenamefont {Gopalakrishna}\ \emph {et~al.}(2008)\citenamefont
  {Gopalakrishna}, \citenamefont {Jung},\ and\ \citenamefont
  {Wells}}]{Gopalakrishna:2008dv}%
  \BibitemOpen
  \bibfield  {author} {\bibinfo {author} {\bibfnamefont {S.}~\bibnamefont
  {Gopalakrishna}}, \bibinfo {author} {\bibfnamefont {S.}~\bibnamefont {Jung}},
  \ and\ \bibinfo {author} {\bibfnamefont {J.~D.}\ \bibnamefont {Wells}},\
  }\Doi {10.1103/PhysRevD.78.055002} {\bibfield  {journal} {\bibinfo  {journal}
  {Phys.Rev.},\ }\textbf {\bibinfo {volume} {D78}},\ \bibinfo {pages} {055002}
  (\bibinfo {year} {2008})},\ \Eprint {http://arxiv.org/abs/0801.3456}
  {arXiv:0801.3456 [hep-ph]} \BibitemShut {NoStop}%
\bibitem [{\citenamefont {Kribs}\ \emph
  {et~al.}(2010){\natexlab{d}}\citenamefont {Kribs}, \citenamefont {Okui},\
  and\ \citenamefont {Roy}}]{Kribs:2010md}%
  \BibitemOpen
  \bibfield  {author} {\bibinfo {author} {\bibfnamefont {G.~D.}\ \bibnamefont
  {Kribs}}, \bibinfo {author} {\bibfnamefont {T.}~\bibnamefont {Okui}}, \ and\
  \bibinfo {author} {\bibfnamefont {T.~S.}\ \bibnamefont {Roy}},\ }\href@noop
  {} { (\bibinfo {year} {2010}{\natexlab{d}})},\ \Eprint
  {http://arxiv.org/abs/1008.1798} {arXiv:1008.1798 [hep-ph]} \BibitemShut
  {NoStop}%
%%CITATION = 1008.1798;%%
\bibitem [{\citenamefont {Gabbiani}\ and\ \citenamefont
  {Masiero}(1989)}]{Gabbiani:1988rb}%
  \BibitemOpen
  \bibfield  {author} {\bibinfo {author} {\bibfnamefont {F.}~\bibnamefont
  {Gabbiani}}\ and\ \bibinfo {author} {\bibfnamefont {A.}~\bibnamefont
  {Masiero}},\ }\Doi {10.1016/0550-3213(89)90492-6} {\bibfield  {journal}
  {\bibinfo  {journal} {Nucl. Phys.},\ }\textbf {\bibinfo {volume} {B322}},\
  \bibinfo {pages} {235} (\bibinfo {year} {1989})}\BibitemShut {NoStop}%
%%CITATION = NUPHA,B322,235;%%
\bibitem [{\citenamefont {Gabbiani}\ \emph {et~al.}(1996)\citenamefont
  {Gabbiani}, \citenamefont {Gabrielli}, \citenamefont {Masiero},\ and\
  \citenamefont {Silvestrini}}]{Gabbiani:1996hi}%
  \BibitemOpen
  \bibfield  {author} {\bibinfo {author} {\bibfnamefont {F.}~\bibnamefont
  {Gabbiani}}, \bibinfo {author} {\bibfnamefont {E.}~\bibnamefont {Gabrielli}},
  \bibinfo {author} {\bibfnamefont {A.}~\bibnamefont {Masiero}}, \ and\
  \bibinfo {author} {\bibfnamefont {L.}~\bibnamefont {Silvestrini}},\ }\Doi
  {10.1016/0550-3213(96)00390-2} {\bibfield  {journal} {\bibinfo  {journal}
  {Nucl. Phys.},\ }\textbf {\bibinfo {volume} {B477}},\ \bibinfo {pages} {321}
  (\bibinfo {year} {1996})},\ \Eprint {http://arxiv.org/abs/hep-ph/9604387}
  {arXiv:hep-ph/9604387} \BibitemShut {NoStop}%
%%CITATION = HEP-PH/9604387;%%
\bibitem [{\citenamefont {Bagger}\ \emph {et~al.}(1997)\citenamefont {Bagger},
  \citenamefont {Matchev},\ and\ \citenamefont {Zhang}}]{Bagger:1997gg}%
  \BibitemOpen
  \bibfield  {author} {\bibinfo {author} {\bibfnamefont {J.~A.}\ \bibnamefont
  {Bagger}}, \bibinfo {author} {\bibfnamefont {K.~T.}\ \bibnamefont {Matchev}},
  \ and\ \bibinfo {author} {\bibfnamefont {R.-J.}\ \bibnamefont {Zhang}},\
  }\Doi {10.1016/S0370-2693(97)00920-9} {\bibfield  {journal} {\bibinfo
  {journal} {Phys. Lett.},\ }\textbf {\bibinfo {volume} {B412}},\ \bibinfo
  {pages} {77} (\bibinfo {year} {1997})},\ \Eprint
  {http://arxiv.org/abs/hep-ph/9707225} {arXiv:hep-ph/9707225} \BibitemShut
  {NoStop}%
%%CITATION = HEP-PH/9707225;%%
\bibitem [{\citenamefont {Ciuchini}\ \emph {et~al.}(1998)\citenamefont
  {Ciuchini} \emph {et~al.}}]{Ciuchini:1998ix}%
  \BibitemOpen
  \bibfield  {author} {\bibinfo {author} {\bibfnamefont {M.}~\bibnamefont
  {Ciuchini}} \emph {et~al.},\ }\href@noop {} {\bibfield  {journal} {\bibinfo
  {journal} {JHEP},\ }\textbf {\bibinfo {volume} {10}},\ \bibinfo {pages} {008}
  (\bibinfo {year} {1998})},\ \Eprint {http://arxiv.org/abs/hep-ph/9808328}
  {arXiv:hep-ph/9808328} \BibitemShut {NoStop}%
%%CITATION = HEP-PH/9808328;%%
\bibitem [{\citenamefont {Chamseddine}\ \emph {et~al.}(1982)\citenamefont
  {Chamseddine}, \citenamefont {Arnowitt},\ and\ \citenamefont
  {Nath}}]{Chamseddine:1982jx}%
  \BibitemOpen
  \bibfield  {author} {\bibinfo {author} {\bibfnamefont {A.~H.}\ \bibnamefont
  {Chamseddine}}, \bibinfo {author} {\bibfnamefont {R.~L.}\ \bibnamefont
  {Arnowitt}}, \ and\ \bibinfo {author} {\bibfnamefont {P.}~\bibnamefont
  {Nath}},\ }\Doi {10.1103/PhysRevLett.49.970} {\bibfield  {journal} {\bibinfo
  {journal} {Phys. Rev. Lett.},\ }\textbf {\bibinfo {volume} {49}},\ \bibinfo
  {pages} {970} (\bibinfo {year} {1982})}\BibitemShut {NoStop}%
%%CITATION = PRLTA,49,970;%%
\bibitem [{\citenamefont {Barbieri}\ \emph {et~al.}(1982)\citenamefont
  {Barbieri}, \citenamefont {Ferrara},\ and\ \citenamefont
  {Savoy}}]{Barbieri:1982eh}%
  \BibitemOpen
  \bibfield  {author} {\bibinfo {author} {\bibfnamefont {R.}~\bibnamefont
  {Barbieri}}, \bibinfo {author} {\bibfnamefont {S.}~\bibnamefont {Ferrara}}, \
  and\ \bibinfo {author} {\bibfnamefont {C.~A.}\ \bibnamefont {Savoy}},\ }\Doi
  {10.1016/0370-2693(82)90685-2} {\bibfield  {journal} {\bibinfo  {journal}
  {Phys. Lett.},\ }\textbf {\bibinfo {volume} {B119}},\ \bibinfo {pages} {343}
  (\bibinfo {year} {1982})}\BibitemShut {NoStop}%
%%CITATION = PHLTA,B119,343;%%
\bibitem [{\citenamefont {Ibanez}(1982)}]{Ibanez:1982ee}%
  \BibitemOpen
  \bibfield  {author} {\bibinfo {author} {\bibfnamefont {L.~E.}\ \bibnamefont
  {Ibanez}},\ }\Doi {10.1016/0370-2693(82)90604-9} {\bibfield  {journal}
  {\bibinfo  {journal} {Phys. Lett.},\ }\textbf {\bibinfo {volume} {B118}},\
  \bibinfo {pages} {73} (\bibinfo {year} {1982})}\BibitemShut {NoStop}%
%%CITATION = PHLTA,B118,73;%%
\bibitem [{\citenamefont {Hall}\ \emph {et~al.}(1983)\citenamefont {Hall},
  \citenamefont {Lykken},\ and\ \citenamefont {Weinberg}}]{Hall:1983iz}%
  \BibitemOpen
  \bibfield  {author} {\bibinfo {author} {\bibfnamefont {L.~J.}\ \bibnamefont
  {Hall}}, \bibinfo {author} {\bibfnamefont {J.~D.}\ \bibnamefont {Lykken}}, \
  and\ \bibinfo {author} {\bibfnamefont {S.}~\bibnamefont {Weinberg}},\ }\Doi
  {10.1103/PhysRevD.27.2359} {\bibfield  {journal} {\bibinfo  {journal} {Phys.
  Rev.},\ }\textbf {\bibinfo {volume} {D27}},\ \bibinfo {pages} {2359}
  (\bibinfo {year} {1983})}\BibitemShut {NoStop}%
%%CITATION = PHRVA,D27,2359;%%
\bibitem [{\citenamefont {Ohta}(1983)}]{Ohta:1982wn}%
  \BibitemOpen
  \bibfield  {author} {\bibinfo {author} {\bibfnamefont {N.}~\bibnamefont
  {Ohta}},\ }\Doi {10.1143/PTP.70.542} {\bibfield  {journal} {\bibinfo
  {journal} {Prog. Theor. Phys.},\ }\textbf {\bibinfo {volume} {70}},\ \bibinfo
  {pages} {542} (\bibinfo {year} {1983})}\BibitemShut {NoStop}%
%%CITATION = PTPKA,70,542;%%
\bibitem [{\citenamefont {Ellis}\ \emph {et~al.}(1983)\citenamefont {Ellis},
  \citenamefont {Nanopoulos},\ and\ \citenamefont {Tamvakis}}]{Ellis:1982wr}%
  \BibitemOpen
  \bibfield  {author} {\bibinfo {author} {\bibfnamefont {J.~R.}\ \bibnamefont
  {Ellis}}, \bibinfo {author} {\bibfnamefont {D.~V.}\ \bibnamefont
  {Nanopoulos}}, \ and\ \bibinfo {author} {\bibfnamefont {K.}~\bibnamefont
  {Tamvakis}},\ }\Doi {10.1016/0370-2693(83)90900-0} {\bibfield  {journal}
  {\bibinfo  {journal} {Phys. Lett.},\ }\textbf {\bibinfo {volume} {B121}},\
  \bibinfo {pages} {123} (\bibinfo {year} {1983})}\BibitemShut {NoStop}%
%%CITATION = PHLTA,B121,123;%%
\bibitem [{\citenamefont {Alvarez-Gaume}\ \emph {et~al.}(1983)\citenamefont
  {Alvarez-Gaume}, \citenamefont {Polchinski},\ and\ \citenamefont
  {Wise}}]{AlvarezGaume:1983gj}%
  \BibitemOpen
  \bibfield  {author} {\bibinfo {author} {\bibfnamefont {L.}~\bibnamefont
  {Alvarez-Gaume}}, \bibinfo {author} {\bibfnamefont {J.}~\bibnamefont
  {Polchinski}}, \ and\ \bibinfo {author} {\bibfnamefont {M.~B.}\ \bibnamefont
  {Wise}},\ }\Doi {10.1016/0550-3213(83)90591-6} {\bibfield  {journal}
  {\bibinfo  {journal} {Nucl. Phys.},\ }\textbf {\bibinfo {volume} {B221}},\
  \bibinfo {pages} {495} (\bibinfo {year} {1983})}\BibitemShut {NoStop}%
%%CITATION = NUPHA,B221,495;%%
\bibitem [{\citenamefont {Nilles}(1984)}]{Nilles:1983ge}%
  \BibitemOpen
  \bibfield  {author} {\bibinfo {author} {\bibfnamefont {H.~P.}\ \bibnamefont
  {Nilles}},\ }\Doi {10.1016/0370-1573(84)90008-5} {\bibfield  {journal}
  {\bibinfo  {journal} {Phys. Rept.},\ }\textbf {\bibinfo {volume} {110}},\
  \bibinfo {pages} {1} (\bibinfo {year} {1984})}\BibitemShut {NoStop}%
%%CITATION = PRPLC,110,1;%%
\bibitem [{\citenamefont {Nath}\ \emph {et~al.}(1984)\citenamefont {Nath},
  \citenamefont {Arnowitt},\ and\ \citenamefont {Chamseddine}}]{Nath:1983fp}%
  \BibitemOpen
  \bibfield  {author} {\bibinfo {author} {\bibfnamefont {P.}~\bibnamefont
  {Nath}}, \bibinfo {author} {\bibfnamefont {R.~L.}\ \bibnamefont {Arnowitt}},
  \ and\ \bibinfo {author} {\bibfnamefont {A.~H.}\ \bibnamefont
  {Chamseddine}},\ }\href@noop {} { (\bibinfo {year} {1984})},\ \bibinfo {note}
  {lectures given at Summer Workshop on Particle Physics, Trieste, Italy, Jun
  20 - Jul 29, 1983}\BibitemShut {NoStop}%
\bibitem [{\citenamefont {Barbieri}\ \emph {et~al.}(1996)\citenamefont
  {Barbieri}, \citenamefont {Dvali},\ and\ \citenamefont
  {Hall}}]{Barbieri:1995uv}%
  \BibitemOpen
  \bibfield  {author} {\bibinfo {author} {\bibfnamefont {R.}~\bibnamefont
  {Barbieri}}, \bibinfo {author} {\bibfnamefont {G.~R.}\ \bibnamefont {Dvali}},
  \ and\ \bibinfo {author} {\bibfnamefont {L.~J.}\ \bibnamefont {Hall}},\ }\Doi
  {10.1016/0370-2693(96)00318-8} {\bibfield  {journal} {\bibinfo  {journal}
  {Phys. Lett.},\ }\textbf {\bibinfo {volume} {B377}},\ \bibinfo {pages} {76}
  (\bibinfo {year} {1996})},\ \Eprint {http://arxiv.org/abs/hep-ph/9512388}
  {arXiv:hep-ph/9512388} \BibitemShut {NoStop}%
%%CITATION = HEP-PH/9512388;%%
\bibitem [{\citenamefont {Barbieri}\ \emph {et~al.}(1997)\citenamefont
  {Barbieri}, \citenamefont {Hall}, \citenamefont {Raby},\ and\ \citenamefont
  {Romanino}}]{Barbieri:1996ww}%
  \BibitemOpen
  \bibfield  {author} {\bibinfo {author} {\bibfnamefont {R.}~\bibnamefont
  {Barbieri}}, \bibinfo {author} {\bibfnamefont {L.~J.}\ \bibnamefont {Hall}},
  \bibinfo {author} {\bibfnamefont {S.}~\bibnamefont {Raby}}, \ and\ \bibinfo
  {author} {\bibfnamefont {A.}~\bibnamefont {Romanino}},\ }\Doi
  {10.1016/S0550-3213(97)00134-X} {\bibfield  {journal} {\bibinfo  {journal}
  {Nucl. Phys.},\ }\textbf {\bibinfo {volume} {B493}},\ \bibinfo {pages} {3}
  (\bibinfo {year} {1997})},\ \Eprint {http://arxiv.org/abs/hep-ph/9610449}
  {arXiv:hep-ph/9610449} \BibitemShut {NoStop}%
%%CITATION = HEP-PH/9610449;%%
\bibitem [{\citenamefont {Kaplan}\ and\ \citenamefont
  {Schmaltz}(1994)}]{Kaplan:1993ej}%
  \BibitemOpen
  \bibfield  {author} {\bibinfo {author} {\bibfnamefont {D.~B.}\ \bibnamefont
  {Kaplan}}\ and\ \bibinfo {author} {\bibfnamefont {M.}~\bibnamefont
  {Schmaltz}},\ }\Doi {10.1103/PhysRevD.49.3741} {\bibfield  {journal}
  {\bibinfo  {journal} {Phys. Rev.},\ }\textbf {\bibinfo {volume} {D49}},\
  \bibinfo {pages} {3741} (\bibinfo {year} {1994})},\ \Eprint
  {http://arxiv.org/abs/hep-ph/9311281} {arXiv:hep-ph/9311281} \BibitemShut
  {NoStop}%
%%CITATION = HEP-PH/9311281;%%
\bibitem [{\citenamefont {Arkani-Hamed}\ \emph {et~al.}(1996)\citenamefont
  {Arkani-Hamed}, \citenamefont {Carone}, \citenamefont {Hall},\ and\
  \citenamefont {Murayama}}]{ArkaniHamed:1996xm}%
  \BibitemOpen
  \bibfield  {author} {\bibinfo {author} {\bibfnamefont {N.}~\bibnamefont
  {Arkani-Hamed}}, \bibinfo {author} {\bibfnamefont {C.~D.}\ \bibnamefont
  {Carone}}, \bibinfo {author} {\bibfnamefont {L.~J.}\ \bibnamefont {Hall}}, \
  and\ \bibinfo {author} {\bibfnamefont {H.}~\bibnamefont {Murayama}},\ }\Doi
  {10.1103/PhysRevD.54.7032} {\bibfield  {journal} {\bibinfo  {journal} {Phys.
  Rev.},\ }\textbf {\bibinfo {volume} {D54}},\ \bibinfo {pages} {7032}
  (\bibinfo {year} {1996})},\ \Eprint {http://arxiv.org/abs/hep-ph/9607298}
  {arXiv:hep-ph/9607298} \BibitemShut {NoStop}%
%%CITATION = HEP-PH/9607298;%%
\bibitem [{\citenamefont {Hall}\ and\ \citenamefont
  {Randall}(1990)}]{Hall:1990ac}%
  \BibitemOpen
  \bibfield  {author} {\bibinfo {author} {\bibfnamefont {L.~J.}\ \bibnamefont
  {Hall}}\ and\ \bibinfo {author} {\bibfnamefont {L.}~\bibnamefont {Randall}},\
  }\Doi {10.1103/PhysRevLett.65.2939} {\bibfield  {journal} {\bibinfo
  {journal} {Phys. Rev. Lett.},\ }\textbf {\bibinfo {volume} {65}},\ \bibinfo
  {pages} {2939} (\bibinfo {year} {1990})}\BibitemShut {NoStop}%
%%CITATION = PRLTA,65,2939;%%
\bibitem [{\citenamefont {Nir}\ and\ \citenamefont
  {Seiberg}(1993)}]{Nir:1993mx}%
  \BibitemOpen
  \bibfield  {author} {\bibinfo {author} {\bibfnamefont {Y.}~\bibnamefont
  {Nir}}\ and\ \bibinfo {author} {\bibfnamefont {N.}~\bibnamefont {Seiberg}},\
  }\Doi {10.1016/0370-2693(93)90942-B} {\bibfield  {journal} {\bibinfo
  {journal} {Phys. Lett.},\ }\textbf {\bibinfo {volume} {B309}},\ \bibinfo
  {pages} {337} (\bibinfo {year} {1993})},\ \Eprint
  {http://arxiv.org/abs/hep-ph/9304307} {arXiv:hep-ph/9304307} \BibitemShut
  {NoStop}%
%%CITATION = HEP-PH/9304307;%%
\bibitem [{\citenamefont {Alwall}\ \emph {et~al.}(2009)\citenamefont {Alwall},
  \citenamefont {Schuster},\ and\ \citenamefont {Toro}}]{Alwall:2008ag}%
  \BibitemOpen
  \bibfield  {author} {\bibinfo {author} {\bibfnamefont {J.}~\bibnamefont
  {Alwall}}, \bibinfo {author} {\bibfnamefont {P.}~\bibnamefont {Schuster}}, \
  and\ \bibinfo {author} {\bibfnamefont {N.}~\bibnamefont {Toro}},\ }\Doi
  {10.1103/PhysRevD.79.075020} {\bibfield  {journal} {\bibinfo  {journal}
  {Phys.Rev.},\ }\textbf {\bibinfo {volume} {D79}},\ \bibinfo {pages} {075020}
  (\bibinfo {year} {2009})},\ \Eprint {http://arxiv.org/abs/0810.3921}
  {arXiv:0810.3921 [hep-ph]} \BibitemShut {NoStop}%
\bibitem [{\citenamefont {Hook}\ \emph {et~al.}(2010)\citenamefont {Hook},
  \citenamefont {Izaguirre},\ and\ \citenamefont {Wacker}}]{Hook:2010tw}%
  \BibitemOpen
  \bibfield  {author} {\bibinfo {author} {\bibfnamefont {A.}~\bibnamefont
  {Hook}}, \bibinfo {author} {\bibfnamefont {E.}~\bibnamefont {Izaguirre}}, \
  and\ \bibinfo {author} {\bibfnamefont {J.~G.}\ \bibnamefont {Wacker}},\
  }\href@noop {} { (\bibinfo {year} {2010})},\ \Eprint
  {http://arxiv.org/abs/1006.0973} {arXiv:1006.0973 [hep-ph]} \BibitemShut
  {NoStop}%
%%CITATION = 1006.0973;%%
\bibitem [{\citenamefont {Djouadi}(2008)}]{Djouadi:2005gj}%
  \BibitemOpen
  \bibfield  {author} {\bibinfo {author} {\bibfnamefont {A.}~\bibnamefont
  {Djouadi}},\ }\Doi {10.1016/j.physrep.2007.10.005} {\bibfield  {journal}
  {\bibinfo  {journal} {Phys.Rept.},\ }\textbf {\bibinfo {volume} {459}},\
  \bibinfo {pages} {1} (\bibinfo {year} {2008})},\ \Eprint
  {http://arxiv.org/abs/hep-ph/0503173} {arXiv:hep-ph/0503173 [hep-ph]}
  \BibitemShut {NoStop}%
\bibitem [{\citenamefont {CDF}(2010)}]{CDF:10238}%
  \BibitemOpen
  \bibfield  {author} {\bibinfo {author} {\bibnamefont {CDF}},\ }\href@noop {}
  {}\bibinfo {type} {Tech. Rep.}\ \bibinfo {number}
  {CDF/PUB/ELECTROWEAK/PUBLIC/10238}\ (\bibinfo {year} {2010})\BibitemShut
  {NoStop}%
\bibitem [{\citenamefont {Alwall}\ \emph {et~al.}(2007)\citenamefont {Alwall}
  \emph {et~al.}}]{Alwall:2007st}%
  \BibitemOpen
  \bibfield  {author} {\bibinfo {author} {\bibfnamefont {J.}~\bibnamefont
  {Alwall}} \emph {et~al.},\ }\Doi {10.1088/1126-6708/2007/09/028} {\bibfield
  {journal} {\bibinfo  {journal} {JHEP},\ }\textbf {\bibinfo {volume} {09}},\
  \bibinfo {pages} {028} (\bibinfo {year} {2007})},\ \Eprint
  {http://arxiv.org/abs/0706.2334} {arXiv:0706.2334 [hep-ph]} \BibitemShut
  {NoStop}%
%%CITATION = 0706.2334;%%
\bibitem [{\citenamefont {Sjostrand}\ \emph {et~al.}(2006)\citenamefont
  {Sjostrand}, \citenamefont {Mrenna},\ and\ \citenamefont
  {Skands}}]{Sjostrand:2006za}%
  \BibitemOpen
  \bibfield  {author} {\bibinfo {author} {\bibfnamefont {T.}~\bibnamefont
  {Sjostrand}}, \bibinfo {author} {\bibfnamefont {S.}~\bibnamefont {Mrenna}}, \
  and\ \bibinfo {author} {\bibfnamefont {P.~Z.}\ \bibnamefont {Skands}},\ }\Doi
  {10.1088/1126-6708/2006/05/026} {\bibfield  {journal} {\bibinfo  {journal}
  {JHEP},\ }\textbf {\bibinfo {volume} {05}},\ \bibinfo {pages} {026} (\bibinfo
  {year} {2006})},\ \Eprint {http://arxiv.org/abs/hep-ph/0603175}
  {arXiv:hep-ph/0603175} \BibitemShut {NoStop}%
%%CITATION = HEP-PH/0603175;%%
\bibitem [{\citenamefont {et~al}()}]{PGS}%
  \BibitemOpen
  \bibfield  {author} {\bibinfo {author} {\bibfnamefont {J.~C.}\ \bibnamefont
  {et~al}},\ }\href@noop {} {\enquote {\bibinfo {title} {{PGS: Pretty Good
  Simulation of high energy collisions}},}\ }\BibitemShut {NoStop}%
\bibitem [{\citenamefont {Anastasiou}\ \emph {et~al.}(2009)\citenamefont
  {Anastasiou}, \citenamefont {Boughezal},\ and\ \citenamefont
  {Petriello}}]{Anastasiou:2008tj}%
  \BibitemOpen
  \bibfield  {author} {\bibinfo {author} {\bibfnamefont {C.}~\bibnamefont
  {Anastasiou}}, \bibinfo {author} {\bibfnamefont {R.}~\bibnamefont
  {Boughezal}}, \ and\ \bibinfo {author} {\bibfnamefont {F.}~\bibnamefont
  {Petriello}},\ }\Doi {10.1088/1126-6708/2009/04/003} {\bibfield  {journal}
  {\bibinfo  {journal} {JHEP},\ }\textbf {\bibinfo {volume} {0904}},\ \bibinfo
  {pages} {003} (\bibinfo {year} {2009})},\ \Eprint
  {http://arxiv.org/abs/0811.3458} {arXiv:0811.3458 [hep-ph]} \BibitemShut
  {NoStop}%
\bibitem [{\citenamefont {Dittmaier}\ \emph {et~al.}(2011)\citenamefont
  {Dittmaier} \emph {et~al.}}]{Dittmaier:2011ti}%
  \BibitemOpen
  \bibfield  {author} {\bibinfo {author} {\bibfnamefont {S.}~\bibnamefont
  {Dittmaier}} \emph {et~al.} (\bibinfo {collaboration} {LHC Higgs Cross
  Section Working Group}),\ }\href@noop {} { (\bibinfo {year} {2011})},\
  \Eprint {http://arxiv.org/abs/1101.0593} {arXiv:1101.0593 [hep-ph]}
  \BibitemShut {NoStop}%
%%CITATION = 1101.0593;%%
\bibitem [{\citenamefont {De~Simone}\ \emph {et~al.}(2008)\citenamefont
  {De~Simone}, \citenamefont {Fan}, \citenamefont {Schmaltz},\ and\
  \citenamefont {Skiba}}]{DeSimone:2008gm}%
  \BibitemOpen
  \bibfield  {author} {\bibinfo {author} {\bibfnamefont {A.}~\bibnamefont
  {De~Simone}}, \bibinfo {author} {\bibfnamefont {J.}~\bibnamefont {Fan}},
  \bibinfo {author} {\bibfnamefont {M.}~\bibnamefont {Schmaltz}}, \ and\
  \bibinfo {author} {\bibfnamefont {W.}~\bibnamefont {Skiba}},\ }\Doi
  {10.1103/PhysRevD.78.095010} {\bibfield  {journal} {\bibinfo  {journal}
  {Phys.Rev.},\ }\textbf {\bibinfo {volume} {D78}},\ \bibinfo {pages} {095010}
  (\bibinfo {year} {2008})},\ \Eprint {http://arxiv.org/abs/0808.2052}
  {arXiv:0808.2052 [hep-ph]} \BibitemShut {NoStop}%
\bibitem [{\citenamefont {De~Simone}\ \emph {et~al.}(2009)\citenamefont
  {De~Simone}, \citenamefont {Fan}, \citenamefont {Sanz},\ and\ \citenamefont
  {Skiba}}]{DeSimone:2009ws}%
  \BibitemOpen
  \bibfield  {author} {\bibinfo {author} {\bibfnamefont {A.}~\bibnamefont
  {De~Simone}}, \bibinfo {author} {\bibfnamefont {J.}~\bibnamefont {Fan}},
  \bibinfo {author} {\bibfnamefont {V.}~\bibnamefont {Sanz}}, \ and\ \bibinfo
  {author} {\bibfnamefont {W.}~\bibnamefont {Skiba}},\ }\Doi
  {10.1103/PhysRevD.80.035010} {\bibfield  {journal} {\bibinfo  {journal}
  {Phys. Rev.},\ }\textbf {\bibinfo {volume} {D80}},\ \bibinfo {pages} {035010}
  (\bibinfo {year} {2009})},\ \Eprint {http://arxiv.org/abs/0903.5305}
  {arXiv:0903.5305 [hep-ph]} \BibitemShut {NoStop}%
%%CITATION = 0903.5305;%%
\bibitem [{\citenamefont {Katz}\ and\ \citenamefont
  {Tweedie}(2010){\natexlab{a}}}]{Katz:2009qx}%
  \BibitemOpen
  \bibfield  {author} {\bibinfo {author} {\bibfnamefont {A.}~\bibnamefont
  {Katz}}\ and\ \bibinfo {author} {\bibfnamefont {B.}~\bibnamefont {Tweedie}},\
  }\Doi {10.1103/PhysRevD.81.035012} {\bibfield  {journal} {\bibinfo  {journal}
  {Phys.Rev.},\ }\textbf {\bibinfo {volume} {D81}},\ \bibinfo {pages} {035012}
  (\bibinfo {year} {2010}{\natexlab{a}})},\ \Eprint
  {http://arxiv.org/abs/0911.4132} {arXiv:0911.4132 [hep-ph]} \BibitemShut
  {NoStop}%
\bibitem [{\citenamefont {Katz}\ and\ \citenamefont
  {Tweedie}(2010){\natexlab{b}}}]{Katz:2010xg}%
  \BibitemOpen
  \bibfield  {author} {\bibinfo {author} {\bibfnamefont {A.}~\bibnamefont
  {Katz}}\ and\ \bibinfo {author} {\bibfnamefont {B.}~\bibnamefont {Tweedie}},\
  }\Doi {10.1103/PhysRevD.81.115003} {\bibfield  {journal} {\bibinfo  {journal}
  {Phys.Rev.},\ }\textbf {\bibinfo {volume} {D81}},\ \bibinfo {pages} {115003}
  (\bibinfo {year} {2010}{\natexlab{b}})},\ \Eprint
  {http://arxiv.org/abs/1003.5664} {arXiv:1003.5664 [hep-ph]} \BibitemShut
  {NoStop}%
\bibitem [{\citenamefont {CDF}(2009)}]{CDF:9817}%
  \BibitemOpen
  \bibfield  {author} {\bibinfo {author} {\bibnamefont {CDF}},\ }\href@noop {}
  {}\bibinfo {type} {Tech. Rep.}\ \bibinfo {number}
  {CDF/PUB/EXOTIC/PUBLIC/9817}\ (\bibinfo {year} {2009})\BibitemShut {NoStop}%
\bibitem [{\citenamefont {Lester}\ and\ \citenamefont
  {Summers}(1999)}]{Lester:1999tx}%
  \BibitemOpen
  \bibfield  {author} {\bibinfo {author} {\bibfnamefont {C.~G.}\ \bibnamefont
  {Lester}}\ and\ \bibinfo {author} {\bibfnamefont {D.~J.}\ \bibnamefont
  {Summers}},\ }\Doi {10.1016/S0370-2693(99)00945-4} {\bibfield  {journal}
  {\bibinfo  {journal} {Phys. Lett.},\ }\textbf {\bibinfo {volume} {B463}},\
  \bibinfo {pages} {99} (\bibinfo {year} {1999})},\ \Eprint
  {http://arxiv.org/abs/hep-ph/9906349} {arXiv:hep-ph/9906349} \BibitemShut
  {NoStop}%
%%CITATION = HEP-PH/9906349;%%
\bibitem [{\citenamefont {Barr}\ \emph {et~al.}(2003)\citenamefont {Barr},
  \citenamefont {Lester},\ and\ \citenamefont {Stephens}}]{Barr:2003rg}%
  \BibitemOpen
  \bibfield  {author} {\bibinfo {author} {\bibfnamefont {A.}~\bibnamefont
  {Barr}}, \bibinfo {author} {\bibfnamefont {C.}~\bibnamefont {Lester}}, \ and\
  \bibinfo {author} {\bibfnamefont {P.}~\bibnamefont {Stephens}},\ }\Doi
  {10.1088/0954-3899/29/10/304} {\bibfield  {journal} {\bibinfo  {journal}
  {J.Phys.G},\ }\textbf {\bibinfo {volume} {G29}},\ \bibinfo {pages} {2343}
  (\bibinfo {year} {2003})},\ \Eprint {http://arxiv.org/abs/hep-ph/0304226}
  {arXiv:hep-ph/0304226 [hep-ph]} \BibitemShut {NoStop}%
\bibitem [{\citenamefont {Cheng}\ and\ \citenamefont
  {Han}(2008)}]{Cheng:2008hk}%
  \BibitemOpen
  \bibfield  {author} {\bibinfo {author} {\bibfnamefont {H.-C.}\ \bibnamefont
  {Cheng}}\ and\ \bibinfo {author} {\bibfnamefont {Z.}~\bibnamefont {Han}},\
  }\Doi {10.1088/1126-6708/2008/12/063} {\bibfield  {journal} {\bibinfo
  {journal} {JHEP},\ }\textbf {\bibinfo {volume} {0812}},\ \bibinfo {pages}
  {063} (\bibinfo {year} {2008})},\ \Eprint {http://arxiv.org/abs/0810.5178}
  {arXiv:0810.5178 [hep-ph]} \BibitemShut {NoStop}%
\bibitem [{\citenamefont {Aad}\ \emph {et~al.}(2009){\natexlab{b}}\citenamefont
  {Aad} \emph {et~al.}}]{Aad:1175196}%
  \BibitemOpen
  \bibfield  {author} {\bibinfo {author} {\bibfnamefont {G.}~\bibnamefont
  {Aad}} \emph {et~al.},\ }\href@noop {} {}\bibinfo {type} {Tech. Rep.}\
  \bibinfo {number} {ATL-PHYS-PUB-2009-082}\ (\bibinfo  {institution} {CERN},\
  \bibinfo {address} {Geneva},\ \bibinfo {year} {2009})\BibitemShut {NoStop}%
\bibitem [{\citenamefont {Polchinski}\ and\ \citenamefont
  {Susskind}(1982)}]{Polchinski:1982an}%
  \BibitemOpen
  \bibfield  {author} {\bibinfo {author} {\bibfnamefont {J.}~\bibnamefont
  {Polchinski}}\ and\ \bibinfo {author} {\bibfnamefont {L.}~\bibnamefont
  {Susskind}},\ }\Doi {10.1103/PhysRevD.26.3661} {\bibfield  {journal}
  {\bibinfo  {journal} {Phys. Rev.},\ }\textbf {\bibinfo {volume} {D26}},\
  \bibinfo {pages} {3661} (\bibinfo {year} {1982})}\BibitemShut {NoStop}%
%%CITATION = PHRVA,D26,3661;%%
\bibitem [{\citenamefont {Hall}\ and\ \citenamefont
  {Randall}(1991)}]{Hall:1990hq}%
  \BibitemOpen
  \bibfield  {author} {\bibinfo {author} {\bibfnamefont {L.~J.}\ \bibnamefont
  {Hall}}\ and\ \bibinfo {author} {\bibfnamefont {L.}~\bibnamefont {Randall}},\
  }\Doi {10.1016/0550-3213(91)90444-3} {\bibfield  {journal} {\bibinfo
  {journal} {Nucl. Phys.},\ }\textbf {\bibinfo {volume} {B352}},\ \bibinfo
  {pages} {289} (\bibinfo {year} {1991})}\BibitemShut {NoStop}%
%%CITATION = NUPHA,B352,289;%%
\bibitem [{\citenamefont {Randall}\ and\ \citenamefont
  {Rius}(1992)}]{Randall:1992cq}%
  \BibitemOpen
  \bibfield  {author} {\bibinfo {author} {\bibfnamefont {L.}~\bibnamefont
  {Randall}}\ and\ \bibinfo {author} {\bibfnamefont {N.}~\bibnamefont {Rius}},\
  }\Doi {10.1016/0370-2693(92)91779-9} {\bibfield  {journal} {\bibinfo
  {journal} {Phys. Lett.},\ }\textbf {\bibinfo {volume} {B286}},\ \bibinfo
  {pages} {299} (\bibinfo {year} {1992})}\BibitemShut {NoStop}%
%%CITATION = PHLTA,B286,299;%%
\bibitem [{\citenamefont {Fox}\ \emph {et~al.}(2002)\citenamefont {Fox},
  \citenamefont {Nelson},\ and\ \citenamefont {Weiner}}]{Fox:2002bu}%
  \BibitemOpen
  \bibfield  {author} {\bibinfo {author} {\bibfnamefont {P.~J.}\ \bibnamefont
  {Fox}}, \bibinfo {author} {\bibfnamefont {A.~E.}\ \bibnamefont {Nelson}}, \
  and\ \bibinfo {author} {\bibfnamefont {N.}~\bibnamefont {Weiner}},\
  }\href@noop {} {\bibfield  {journal} {\bibinfo  {journal} {JHEP},\ }\textbf
  {\bibinfo {volume} {08}},\ \bibinfo {pages} {035} (\bibinfo {year} {2002})},\
  \Eprint {http://arxiv.org/abs/hep-ph/0206096} {arXiv:hep-ph/0206096}
  \BibitemShut {NoStop}%
%%CITATION = HEP-PH/0206096;%%
\bibitem [{\citenamefont {Nelson}\ \emph {et~al.}(2002)\citenamefont {Nelson},
  \citenamefont {Rius}, \citenamefont {Sanz},\ and\ \citenamefont
  {Unsal}}]{Nelson:2002ca}%
  \BibitemOpen
  \bibfield  {author} {\bibinfo {author} {\bibfnamefont {A.~E.}\ \bibnamefont
  {Nelson}}, \bibinfo {author} {\bibfnamefont {N.}~\bibnamefont {Rius}},
  \bibinfo {author} {\bibfnamefont {V.}~\bibnamefont {Sanz}}, \ and\ \bibinfo
  {author} {\bibfnamefont {M.}~\bibnamefont {Unsal}},\ }\href@noop {}
  {\bibfield  {journal} {\bibinfo  {journal} {JHEP},\ }\textbf {\bibinfo
  {volume} {08}},\ \bibinfo {pages} {039} (\bibinfo {year} {2002})},\ \Eprint
  {http://arxiv.org/abs/hep-ph/0206102} {arXiv:hep-ph/0206102} \BibitemShut
  {NoStop}%
%%CITATION = HEP-PH/0206102;%%
\bibitem [{\citenamefont {Chacko}\ \emph {et~al.}(2005)\citenamefont {Chacko},
  \citenamefont {Fox},\ and\ \citenamefont {Murayama}}]{Chacko:2004mi}%
  \BibitemOpen
  \bibfield  {author} {\bibinfo {author} {\bibfnamefont {Z.}~\bibnamefont
  {Chacko}}, \bibinfo {author} {\bibfnamefont {P.~J.}\ \bibnamefont {Fox}}, \
  and\ \bibinfo {author} {\bibfnamefont {H.}~\bibnamefont {Murayama}},\ }\Doi
  {10.1016/j.nuclphysb.2004.11.021} {\bibfield  {journal} {\bibinfo  {journal}
  {Nucl. Phys.},\ }\textbf {\bibinfo {volume} {B706}},\ \bibinfo {pages} {53}
  (\bibinfo {year} {2005})},\ \Eprint {http://arxiv.org/abs/hep-ph/0406142}
  {arXiv:hep-ph/0406142} \BibitemShut {NoStop}%
%%CITATION = HEP-PH/0406142;%%
\bibitem [{\citenamefont {Carone}\ \emph {et~al.}(2005)\citenamefont {Carone},
  \citenamefont {Erlich},\ and\ \citenamefont {Glover}}]{Carone:2005iq}%
  \BibitemOpen
  \bibfield  {author} {\bibinfo {author} {\bibfnamefont {C.~D.}\ \bibnamefont
  {Carone}}, \bibinfo {author} {\bibfnamefont {J.}~\bibnamefont {Erlich}}, \
  and\ \bibinfo {author} {\bibfnamefont {B.}~\bibnamefont {Glover}},\
  }\href@noop {} {\bibfield  {journal} {\bibinfo  {journal} {JHEP},\ }\textbf
  {\bibinfo {volume} {10}},\ \bibinfo {pages} {042} (\bibinfo {year} {2005})},\
  \Eprint {http://arxiv.org/abs/hep-ph/0509002} {arXiv:hep-ph/0509002}
  \BibitemShut {NoStop}%
%%CITATION = HEP-PH/0509002;%%
\bibitem [{\citenamefont {Nomura}\ \emph {et~al.}(2006)\citenamefont {Nomura},
  \citenamefont {Poland},\ and\ \citenamefont {Tweedie}}]{Nomura:2005rj}%
  \BibitemOpen
  \bibfield  {author} {\bibinfo {author} {\bibfnamefont {Y.}~\bibnamefont
  {Nomura}}, \bibinfo {author} {\bibfnamefont {D.}~\bibnamefont {Poland}}, \
  and\ \bibinfo {author} {\bibfnamefont {B.}~\bibnamefont {Tweedie}},\ }\Doi
  {10.1016/j.nuclphysb.2006.03.034} {\bibfield  {journal} {\bibinfo  {journal}
  {Nucl. Phys.},\ }\textbf {\bibinfo {volume} {B745}},\ \bibinfo {pages} {29}
  (\bibinfo {year} {2006})},\ \Eprint {http://arxiv.org/abs/hep-ph/0509243}
  {arXiv:hep-ph/0509243} \BibitemShut {NoStop}%
%%CITATION = HEP-PH/0509243;%%
\bibitem [{\citenamefont {Nomura}\ and\ \citenamefont
  {Tweedie}(2005)}]{Nomura:2005qg}%
  \BibitemOpen
  \bibfield  {author} {\bibinfo {author} {\bibfnamefont {Y.}~\bibnamefont
  {Nomura}}\ and\ \bibinfo {author} {\bibfnamefont {B.}~\bibnamefont
  {Tweedie}},\ }\Doi {10.1103/PhysRevD.72.015006} {\bibfield  {journal}
  {\bibinfo  {journal} {Phys. Rev.},\ }\textbf {\bibinfo {volume} {D72}},\
  \bibinfo {pages} {015006} (\bibinfo {year} {2005})},\ \Eprint
  {http://arxiv.org/abs/hep-ph/0504246} {arXiv:hep-ph/0504246} \BibitemShut
  {NoStop}%
%%CITATION = HEP-PH/0504246;%%
\bibitem [{\citenamefont {Carpenter}\ \emph {et~al.}(2005)\citenamefont
  {Carpenter}, \citenamefont {Fox},\ and\ \citenamefont
  {Kaplan}}]{Carpenter:2005tz}%
  \BibitemOpen
  \bibfield  {author} {\bibinfo {author} {\bibfnamefont {L.~M.}\ \bibnamefont
  {Carpenter}}, \bibinfo {author} {\bibfnamefont {P.~J.}\ \bibnamefont {Fox}},
  \ and\ \bibinfo {author} {\bibfnamefont {D.~E.}\ \bibnamefont {Kaplan}},\
  }\href@noop {} { (\bibinfo {year} {2005})},\ \Eprint
  {http://arxiv.org/abs/hep-ph/0503093} {arXiv:hep-ph/0503093} \BibitemShut
  {NoStop}%
%%CITATION = HEP-PH/0503093;%%
\bibitem [{\citenamefont {Antoniadis}\ \emph {et~al.}(2008)\citenamefont
  {Antoniadis}, \citenamefont {Benakli}, \citenamefont {Delgado},\ and\
  \citenamefont {Quiros}}]{Antoniadis:2006uj}%
  \BibitemOpen
  \bibfield  {author} {\bibinfo {author} {\bibfnamefont {I.}~\bibnamefont
  {Antoniadis}}, \bibinfo {author} {\bibfnamefont {K.}~\bibnamefont {Benakli}},
  \bibinfo {author} {\bibfnamefont {A.}~\bibnamefont {Delgado}}, \ and\
  \bibinfo {author} {\bibfnamefont {M.}~\bibnamefont {Quiros}},\ }\href@noop {}
  {\bibfield  {journal} {\bibinfo  {journal} {Adv. Stud. Theor. Phys.},\
  }\textbf {\bibinfo {volume} {2}},\ \bibinfo {pages} {645} (\bibinfo {year}
  {2008})},\ \Eprint {http://arxiv.org/abs/hep-ph/0610265}
  {arXiv:hep-ph/0610265} \BibitemShut {NoStop}%
%%CITATION = HEP-PH/0610265;%%
\bibitem [{\citenamefont {Nakayama}\ \emph {et~al.}(2007)\citenamefont
  {Nakayama}, \citenamefont {Taki}, \citenamefont {Watari},\ and\ \citenamefont
  {Yanagida}}]{Nakayama:2007cf}%
  \BibitemOpen
  \bibfield  {author} {\bibinfo {author} {\bibfnamefont {Y.}~\bibnamefont
  {Nakayama}}, \bibinfo {author} {\bibfnamefont {M.}~\bibnamefont {Taki}},
  \bibinfo {author} {\bibfnamefont {T.}~\bibnamefont {Watari}}, \ and\ \bibinfo
  {author} {\bibfnamefont {T.~T.}\ \bibnamefont {Yanagida}},\ }\Doi
  {10.1016/j.physletb.2007.08.064} {\bibfield  {journal} {\bibinfo  {journal}
  {Phys. Lett.},\ }\textbf {\bibinfo {volume} {B655}},\ \bibinfo {pages} {58}
  (\bibinfo {year} {2007})},\ \Eprint {http://arxiv.org/abs/0705.0865}
  {arXiv:0705.0865 [hep-ph]} \BibitemShut {NoStop}%
%%CITATION = 0705.0865;%%
\bibitem [{\citenamefont {Amigo}\ \emph {et~al.}(2009)\citenamefont {Amigo},
  \citenamefont {Blechman}, \citenamefont {Fox},\ and\ \citenamefont
  {Poppitz}}]{Amigo:2008rc}%
  \BibitemOpen
  \bibfield  {author} {\bibinfo {author} {\bibfnamefont {S.~D.~L.}\
  \bibnamefont {Amigo}}, \bibinfo {author} {\bibfnamefont {A.~E.}\ \bibnamefont
  {Blechman}}, \bibinfo {author} {\bibfnamefont {P.~J.}\ \bibnamefont {Fox}}, \
  and\ \bibinfo {author} {\bibfnamefont {E.}~\bibnamefont {Poppitz}},\ }\Doi
  {10.1088/1126-6708/2009/01/018} {\bibfield  {journal} {\bibinfo  {journal}
  {JHEP},\ }\textbf {\bibinfo {volume} {01}},\ \bibinfo {pages} {018} (\bibinfo
  {year} {2009})},\ \Eprint {http://arxiv.org/abs/0809.1112} {arXiv:0809.1112
  [hep-ph]} \BibitemShut {NoStop}%
%%CITATION = 0809.1112;%%
\bibitem [{\citenamefont {Kribs}\ \emph {et~al.}(2008)\citenamefont {Kribs},
  \citenamefont {Poppitz},\ and\ \citenamefont {Weiner}}]{Kribs:2007ac}%
  \BibitemOpen
  \bibfield  {author} {\bibinfo {author} {\bibfnamefont {G.~D.}\ \bibnamefont
  {Kribs}}, \bibinfo {author} {\bibfnamefont {E.}~\bibnamefont {Poppitz}}, \
  and\ \bibinfo {author} {\bibfnamefont {N.}~\bibnamefont {Weiner}},\ }\Doi
  {10.1103/PhysRevD.78.055010} {\bibfield  {journal} {\bibinfo  {journal}
  {Phys. Rev.},\ }\textbf {\bibinfo {volume} {D78}},\ \bibinfo {pages} {055010}
  (\bibinfo {year} {2008})},\ \Eprint {http://arxiv.org/abs/0712.2039}
  {arXiv:0712.2039 [hep-ph]} \BibitemShut {NoStop}%
%%CITATION = 0712.2039;%%
\bibitem [{\citenamefont {Kribs}\ \emph {et~al.}(2009)\citenamefont {Kribs},
  \citenamefont {Martin},\ and\ \citenamefont {Roy}}]{Kribs:2009zy}%
  \BibitemOpen
  \bibfield  {author} {\bibinfo {author} {\bibfnamefont {G.~D.}\ \bibnamefont
  {Kribs}}, \bibinfo {author} {\bibfnamefont {A.}~\bibnamefont {Martin}}, \
  and\ \bibinfo {author} {\bibfnamefont {T.~S.}\ \bibnamefont {Roy}},\ }\Doi
  {10.1088/1126-6708/2009/06/042} {\bibfield  {journal} {\bibinfo  {journal}
  {JHEP},\ }\textbf {\bibinfo {volume} {06}},\ \bibinfo {pages} {042} (\bibinfo
  {year} {2009})},\ \Eprint {http://arxiv.org/abs/0901.4105} {arXiv:0901.4105
  [hep-ph]} \BibitemShut {NoStop}%
%%CITATION = 0901.4105;%%
\bibitem [{\citenamefont {Fok}\ and\ \citenamefont {Kribs}(2010)}]{Fok:2010vk}%
  \BibitemOpen
  \bibfield  {author} {\bibinfo {author} {\bibfnamefont {R.}~\bibnamefont
  {Fok}}\ and\ \bibinfo {author} {\bibfnamefont {G.~D.}\ \bibnamefont
  {Kribs}},\ }\Doi {10.1103/PhysRevD.82.035010} {\bibfield  {journal} {\bibinfo
   {journal} {Phys.Rev.},\ }\textbf {\bibinfo {volume} {D82}},\ \bibinfo
  {pages} {035010} (\bibinfo {year} {2010})},\ \Eprint
  {http://arxiv.org/abs/1004.0556} {arXiv:1004.0556 [hep-ph]} \BibitemShut
  {NoStop}%
\end{thebibliography}%

\end{document}